\documentclass[12pt]{article}%
\usepackage{cite}
\usepackage{graphicx}
\usepackage{multicol}
\usepackage{amsfonts}
\usepackage{amssymb}
\usepackage{amsmath}
\usepackage{heck}
\usepackage{hyperref}
\usepackage{setspace}
\usepackage[all]{xy}
\usepackage{verbatim}
\usepackage{color}
\usepackage{epsfig}
\usepackage{cancel}%
\providecommand{\U}[1]{\protect\rule{.1in}{.1in}}
\numberwithin{equation}{section}

\newcommand{\cO}{\mathcal{O}}

\newcommand{\Tr}{\, {\rm Tr}}

\begin{document}

\date{August 2011}

\preprint{MCTP-11-29}

\title{Electroweak Symmetry Breaking in the DSSM}

\institution{IAS}{\centerline{${}^{1}$School of Natural Sciences, Institute for Advanced Study, Princeton, NJ 08540, USA}}

\institution{COLUMBIA}{\centerline{${}^{2}$Department of Physics \& ISCAP, Columbia University, New York, NY 10027, USA}}

\institution{HARVARD}{\centerline{${}^{3}$Jefferson Physical Laboratory, Harvard University, Cambridge, MA 02138, USA}}

\institution{MICH}{\centerline{${}^4$Michigan Center for Theoretical Physics, University of Michigan, Ann Arbor, MI 48109, USA}}

\authors{Jonathan J. Heckman\worksat{\IAS}\footnote{e-mail: {\tt jheckman@ias.edu}}, Piyush Kumar \worksat{\COLUMBIA}\footnote{e-mail: {\tt kpiyush@phys.columbia.edu}}, \\[4mm] Cumrun Vafa\worksat{\HARVARD}\footnote{e-mail: {\tt vafa@physics.harvard.edu}}, and Brian Wecht \worksat{\MICH}\footnote{e-mail: {\tt bwecht@umich.edu}}}

\abstract{We study the theoretical and phenomenological consequences of modifying the K\"ahler potential of the
MSSM two Higgs doublet sector. Such modifications
naturally arise when the Higgs sector mixes with a quasi-hidden conformal sector, as in some F-theory GUT models.
In the Delta-deformed Supersymmetric Standard Model (DSSM), the Higgs fields are
operators with non-trivial scaling dimension $1 < \Delta < 2$. The K\"ahler metric
is singular at the origin of field space due to the presence of quasi-hidden sector states which
get their mass from the Higgs vevs. The presence of these extra states leads to the fact that even
as $\Delta \rightarrow 1$, the DSSM does not reduce to the MSSM. In particular, the Higgs can naturally be heavier than
the $W$- and $Z$-bosons. Perturbative gauge coupling unification,
a large top quark Yukawa, and consistency with precision electroweak
can all be maintained for $\Delta$ close to unity. Moreover, such values
of $\Delta$ can naturally be obtained in string-motivated constructions. The
quasi-hidden sector generically contains states charged under $SU(5)_{GUT}$ as well
as gauge singlets, leading to a rich, albeit model-dependent, collider phenomenology.
}

\maketitle

\enlargethispage{\baselineskip}

\setcounter{tocdepth}{2}
\tableofcontents

\newpage

\section{Introduction}

The origin of electroweak symmetry breaking remains poorly understood. Though
low energy supersymmetry provides a promising framework for addressing the
hierarchy problem, the absence of any evidence from LEP, the Tevatron, and the LHC
already places strong constraints on many supersymmetric models. For example,
pushing the superpartner masses up in the Minimal Supersymmetric Standard
Model (MSSM) leads to a level of fine tuning which is on the order of (at
least) $1\%-0.1\%$ in many models.

From a top-down perspective, however, the MSSM can be viewed as one very
particular supersymmetric field theory. Various SUSY-preserving or
SUSY-breaking deformations of this theoretical structure are in principle possible.
For example, two basic inputs in a supersymmetric field theory are the
K\"{a}hler potential and the superpotential, both of which can be supplemented in
phenomenological models by SUSY-breaking terms. While holomorphy
considerations constrain the superpotential, the K\"ahler potential is far
less constrained. In this regard, the assumption of a canonical K\"ahler
potential in the MSSM is a rather special choice. Although it is the unique
renormalizable choice, in the spirit of effective field theory, more
general choices are {\it a priori} equally plausible.

In this paper, we study the consequences of giving the Higgs sector a K\"{a}hler metric which is singular at the origin of field space.
This can occur when the Higgs fields couple to a quasi-hidden sector in which at least some of the
states of the hidden sector get their mass from the Higgs vevs. From a top-down perspective, this sort of visible sector/hidden sector mixing is quite
well-motivated. For example, in GUT models arising within F-theory, the Standard Model is
realized on a stack of branes, and natural hidden sectors arise when
additional branes are in the vicinity of the Standard Model stack. This leads
to additional states charged under $G_{SM}=SU(3)\times SU(2)\times U(1)$
which mix with the Standard Model. The hidden sector has $SU(5)_{GUT}$ as a weakly gauged flavor
symmetry. In many well-motivated examples, the states fill out vector-like pairs of
representations of $SU(5)_{GUT}$. However, they are assumed to get their mass predominantly from Higgs vevs. Within string theory, this can happen naturally.\footnote{The reason these states remain light in the string theory context is that the bare mass of these vector-like states is controlled by the position of the D3-brane relative to the Standard Model branes. Fluxes which attract the D3-brane to
the SM\ branes thus lead to massless states. See \cite{Funparticles, FCFT,
TBRANES, D3gen, HVW} for recent work on these quasi-hidden sectors. Let us note that in previous studies of these quasi-hidden sectors,
the case of a D3-brane slightly displaced (at energy scales far below the GUT scale) from the Standard Model stack was mainly treated. In this paper we instead focus on the limit where this displacement vanishes and the Higgs vevs are the main source of CFT breaking effects.}
Both of these features will turn out to be relevant for phenomenology.

In contrast to much of the model-building literature, these systems provide
concrete $\mathcal{N}=1$ superconformal field theories which are
\textit{different} from (conformal) SQCD. In these D3-brane CFTs,
the hidden sector does not induce a Landau pole, and can actually improve
precision unification \cite{HVW}. Further, the contribution to the anomalous
dimensions of Standard Model fields is small, allowing these states to retain
their weakly coupled identity \cite{HVW}. Finally, since these CFTs involve one, or at most
two D3-branes, they do not have a conventional AdS dual, which would have involved a large number of
D3-branes.

We can state the main features of the system we shall be studying in purely field
theoretic terms, as we will not need many explicit features of the string constructions. The
reader interested in string theory details is encouraged to refer to Appendix \ref{d3cft} and references therein.
In the UV, the Higgs sector mixes with the hidden sector via:%
\begin{equation}
\delta L_{mix}=\int d^{2}\theta\,\,\left ( H_{u}{\cal O}_{u}+H_{d}{\cal O}_{d} \right ) + h.c. \label{LMIXER}%
\end{equation}
where the ${\cal O}$'s are operators in some hidden sector. Scenarios of electroweak symmetry breaking
involving such mixing terms have been considered previously in
\cite{Stancato:2008mp, Azatov:2011ht, Azatov:2011ps, Gherghetta:2011na}.
When there is non-trivial mixing between the two sectors,
we expect that at lower scales there is an approximately conformal phase where $H_{u}$
and $H_{d}$ pick up non-trivial scaling dimensions $\Delta_{u}$ and
$\Delta_{d}$. We refer to this deformation of the MSSM as the Delta-deformed
Supersymmetric Standard Model (DSSM). Our main interest will be in the regime
where the excess Higgs dimension $\Delta - 1$ is small.

We assume that at a scale $\Lambda_{soft}$, the Higgs sector is deformed by the analogue of
the soft breaking parameters in the usual MSSM:\footnote{Throughout this work, we
shall make the assumption that the dimensions of operators are additive, even
for non-chiral combinations. This is not true for non-chiral
operators, but we expect this is a subleading effect, at least in the limit
where $\Delta\rightarrow1$. Away from this limit, it is a simplifying
assumption we make. See \cite{Rattazzi:2008pe, Poland:2010wg,
Fitzpatrick:2011hh} for recent discussion on this issue.}
\begin{equation}
\delta L_{Higgs}=\left (\int d^{2}\theta\text{ }\mu H_{u}H_{d}  + h.c. \right ) +(BH_{u}H_{d}%
+h.c.)+m_{u}^{2}\left\Vert H_{u}\right\Vert ^{2}+m_{d}^{2}\left\Vert
H_{d}\right\Vert ^{2}.%
\end{equation}
The full details of the Higgs potential are then fixed by the K\"ahler potential
$K$. We do not know the precise form of $K$, but using simple scaling arguments, it is possible to deduce some
generic features of the DSSM Higgs potential. In the DSSM Higgs potential, the presence of fractional powers of fields in the
K\"ahler metric leads to a competition between at least two
terms in the Higgs potential of degree close to two, such as $\vert H \vert^{2 + 2 \delta}$ and $\vert H \vert^{2 - 2 \delta}$ where $\delta = \Delta - 1$ is the excess Higgs dimensions. This leads to a squeezed Mexican hat potential, in which the vev is naturally
much smaller than the Higgs mass. We find using general arguments that the lightest
Higgs mass $M_{H}$ is related to the gauge boson mass $M_{V}$ by:
\begin{equation}
\frac{M_{H}}{M_{V}}\sim\frac{\sqrt{\delta}}{g}\times(\sqrt{q_{0}}%
)^{1/\delta},%
\end{equation}
where $g$ is a gauge coupling, and $q_{0}<1$ is an order one parameter fixed by the details of the Higgs
potential. We also check the form of these expressions for a particular choice of
K\"ahler potential.

As $\Delta\rightarrow1$, the Higgs becomes parametrically heavier than the gauge bosons. This illustrates an important feature
of the DSSM: No matter how close $\Delta$ gets to one, there is an additional minimum in the Higgs potential which is not present in the
case when $\Delta = 1$. It is this feature which eliminates the usual fine tuning present in the Higgs
sector of the MSSM. In the regime $\Delta \rightarrow 1$, the extra states of the DSSM can be viewed very roughly as additional vector-like generations,
but where an effectively non-integer number of fields contribute to the gauge coupling beta functions. As $\Delta
\rightarrow2$, one is far from the regime of weak coupling. In this limit, the
K\"{a}hler potential may include terms such as $\sqrt{H^{\dag}H}$, which is
more in line with a composite Higgs. See for example \cite{Luty:2000fj, Harnik:2003rs, Stancato:2008mp, Stancato:2010ay, Fukushima:2010pm,
Craig:2011ev, Azatov:2011ht, Azatov:2011ps, Gherghetta:2011na} for other recent work on coupling the
Higgs to a strongly coupled hidden sector.

The regime of small $\delta=\Delta-1$ is attractive for a number of reasons.
In this case, the top quark Yukawa can remain reasonably
large, as it is close to being a marginal coupling. Further, when $\delta\leq0.1$,
perturbative gauge coupling unification is retained
\cite{HVW} (see also \cite{Donkin:2010ta}). As we
explain later, the regime $0.01\leq\delta\leq0.1$ is also favored by constraints
from precision electroweak physics. Such values of $\delta$ can also be naturally obtained in explicit models \cite{HVW}.

In summary, the characteristic mass scales of the DSSM are:%
\begin{align}
M_{V}  &  \sim\Lambda_{soft}\times g\times(\sqrt{q_{0}})^{1/\delta}\\
M_{H}  &  \sim\Lambda_{soft}\times\sqrt{\delta}\\
M_{extra}  &  \sim4\pi v\times\sqrt{\delta}\sim(3\text{ TeV})\times
\sqrt{\delta}%
\end{align}
where $M_{extra}$ is the mass of the extra states which get their mass from
the Higgs vev, and lead to the singular behavior in the K\"{a}hler potential.

The presence of these extra states is motivated by UV considerations, but has
direct consequences for experiment. Indeed, since these extra states couple to
the Higgs, they will be charged under $G_{SM}$. Another common feature of these quasi-hidden
sectors is the presence of additional SM\ gauge singlets which participate in the strongly coupled hidden sector.
The signatures of this framework have considerable overlap with a particular class of hidden-valley models
\cite{Strassler:2006im, Strassler:2008bv}. See also \cite{Cacciapaglia:2007jq, Cacciapaglia:2008ns, Cai:2009ax, Cai:2011ww} for related work on the phenomenology of unparticle scenarios with a mass gap, and in particular \cite{Stancato:2008mp, Stancato:2010ay} for some discussion of the consequences for Higgs physics.

The rest of this paper is organized as follows. First, in section
\ref{sec:SETUP} we describe the general setup. In section \ref{sec:SPECTRUM}
we study the mass scales of the DSSM. In section \ref{sec:CONSTRAINTS}
we discuss precision electroweak constraints, and in section
\ref{sec:collider} we discuss some qualitative features of the associated
collider phenomenology. Section \ref{sec:CONCLUDE} contains our conclusions.
In the Appendices we present some additional details and explicit examples.

\section{General Setup \label{sec:SETUP}}

In this section we state the general setup for the DSSM. We assume
that in the ultraviolet regime, e.g. the GUT/string scale, the field content
of the Standard Model is weakly coupled to a strongly coupled quasi-hidden
sector. From the perspective of the hidden sector, the Standard Model gauge
group is simply a weakly gauged flavor symmetry, according to which the
operators must be organized. Thus, these operators will in general transform
under some representation of the Standard Model gauge group.

At this high energy scale, we add interaction terms between the two sectors.
For our purposes, these can be summarized as mixing $\cO_{MSSM} \cO_{HID}$
between operators $\cO_{MSSM}$ of the MSSM and $\cO_{HID}$ of the hidden sector.
It is worth noting that in many cases of interest, the hidden sector may not
even possess a Lagrangian formulation. In such cases, the meaning of this
operator deformation is that it is to be inserted into the correlation
functions for the theory. We focus on mixing between the Higgs and hidden
sectors via the F-terms:%
\begin{equation}
\delta L_{mix}=\int d^{2}\theta\text{ }(H_{u} \cO_{u}+H_{d} \cO_{d}) + h.c..
\end{equation}
Such couplings naturally arise in string inspired models where operators from
a D3-brane sector such as $\cO_{u}$ and $\cO_{d}$ couple to Standard Model
operators localized on a seven-branes. Our main assumption
(which can be justified in some string models) is that these
deformations are relevant in the UV theory and cause the Higgs
fields to develop a non-trivial (but not very different from one)
scaling dimension in the IR. See \cite{Stancato:2008mp, Azatov:2011ht, Azatov:2011ps, Gherghetta:2011na} for
recent work on related scenarios of electroweak symmetry breaking involving the effects of such couplings.

In general we can say relatively little when the mixing between the Higgs fields and $\cO$ operators is large.
However, in the limit where the Higgs dimensions stay close
to one, we can exercise more control. This type of behavior can be realized,
as in the examples of \cite{HVW}. Such deformations often trigger a
renormalization group flow to a new interacting conformal fixed point. One way of characterizing this
behavior is that in flowing from the UV to the IR, the fields $H_{u}$ and
$H_{d}$ pick up non-trivial scaling dimensions. Denote by $\Delta_{u}$ and $\Delta_{d}$ the scaling
dimensions of the Higgs fields $H_u$ and $H_d$. It is also helpful to introduce:
\begin{equation}
\Delta \equiv \frac{\Delta_u + \Delta_d}{2}\,\, , \,\, \delta \equiv \Delta -1.
\end{equation}
We shall often use $\Delta$ to characterize the effects of
non-trivial Higgs scaling dimensions. As will become clear, the regime
$\Delta \approx 1$ appears to be most natural. Thus we often limit
our analysis to an expansion in small $\delta$.

Of course, in the real world, we never reach such a fixed point. This is
because supersymmetry is broken at some scale, so conformal symmetry must be broken at this scale or above.
We assume that the energy scale where this
breakdown of conformal symmetry occurs depends on the soft mass scale
$\Lambda_{soft}$. This scale enters through the $\mu$-term, $B\mu$-term, and
the analogue of the soft mass terms, which we assume are all given by powers
of $\Lambda_{soft}$, up to order one coefficients\footnote{In this work, we do not discuss the dynamical mechanism by which supersymmetry is broken. Instead, we assume supersymmetry breaking and discuss the nature of the soft terms consistent with the structure of the framework. As a simplifying assumption, in this work we assume that supersymmetry breaking is decoupled from the dynamics of the quasi-conformal hidden sector. It would be interesting to relax this assumption.}. We now discuss each of these terms.

The $\mu$-term is an F-term and is holomorphic in the chiral operators
$H_{u}$ and $H_{d}$. The superpotential is $\mu H_{u}H_{d}$, so $\mu$ scales as $\mu=\mu_{0}\Lambda
_{soft}^{3-\Delta_{u}-\Delta_{d}}$, with $\mu_{0}$ a dimensionless constant.
Similarly, because the $B\mu$ term involves the descendant of a chiral
operator, we have a term $BH_{u}H_{d}+h.c.$ in the Lagrangian density. Because this term involves
the scalar components of chiral operators, the coefficient $B$ is
fixed as $B=B_{0}\Lambda_{soft}^{4-\Delta_{u}-\Delta_{d}}$, with $B_{0}$ a
dimensionless constant.

The analogue of soft mass terms in the DSSM is more subtle. In the absence of
supersymmetry, it is difficult to constrain the form of such contributions.
However, with supersymmetry, we expect these contributions to be generated
by two-point functions involving the Higgs fields\footnote{For example, this happens in minimal gauge mediation.}.
We shall therefore make the natural assumption that these soft masses are of the form:%
\begin{equation}
m_{u}^{2}\left\Vert H_{u}\right\Vert ^{2}+m_{d}%
^{2}\left\Vert H_{d}\right\Vert ^{2},%
\end{equation}
where the norm appearing in $\left\Vert H\right\Vert ^{2}$ is with
respect to the K\"{a}hler metric, so that $\left\Vert \Phi\right\Vert
^{2}=\Phi^{i}g_{i\overline{j}}\Phi^{\overline{j}\dag}$. Scaling arguments then imply
$m=m_{0}\times\Lambda_{soft}$ for $m_{0}$ a dimensionless number. Note that
the naive engineering dimensions for soft masses work out since the K\"{a}hler potential has
dimension two.

In more formal terms, the full deformation is:\footnote{The more
precise characterization of this deformation is as follows. We suppose that
the Higgs fields interact with a strongly coupled field theory which flows in
the infrared to a strongly coupled conformal theory. As we flow to the
infrared, we suppose that at a scale $\Lambda_{soft}$ this theory is deformed
by an operator $\cO_{deform}$. This means all correlators are to be evaluated
with insertions of $\exp\int d^{4}x$ $\cO_{deform}$. Even the precise definition
of this deformation contains subtleties, because the various terms involve bilinears in
operators of the strongly coupled theory. What is meant here is that all
operators are to be evaluated with respect to the operator product expansion,
with the constant part removed.}%
\begin{equation}\label{deformation}
\cO_{deform}=\left (\int d^{2}\theta\text{ }\mu H_{u}H_{d}  + h.c. \right ) +(BH_{u}H_{d}%
+h.c.)+m_{u}^{2}\left\Vert H_{u}\right\Vert ^{2}+m_{d}^{2}\left\Vert
H_{d}\right\Vert ^{2}.
\end{equation}
The requirement that supersymmetry is softly broken is that the parameters
$B$, $m_{u}^{2}$, and $m_{d}^{2}$ tend to zero as $\Lambda_{soft}\rightarrow0$.
For the soft masses, this does not impose a condition on the dimensions. Note,
however, that vanishing of the $B\mu$ term then requires $\Delta<2$. Although
the $\mu$-term is compatible with supersymmetry, if we also assume that this
vanishes as $\Lambda_{soft}\rightarrow0$, we obtain the sharper bound $\Delta<3/2$. We shall consider the
weaker bound to keep our discussion more general. Finally, we note that we can in principle
also include four-point functions involving the Higgs fields. However, we will find that the terms which are close to degree two
dominate over such contributions, so in what follows we neglect such terms.

Given these operator deformations, we are interested in analyzing the
resulting low energy behavior of the system. By using general scaling arguments, we will show
that after adding these operator deformations, the Higgs develops a vev at some lower scale. We would then like to
determine some features of the low energy spectrum of states. Since there is a
characteristic mass scale for the theory, it makes sense to speak of an
effective potential for the Higgs fields, and in particular, to track the
energy of the vacuum as a function of the Higgs vevs.

\subsection{The K\"{a}hler Potential}\label{sec:Kahlerform}

The precise form of the K\"{a}hler potential clearly plays a crucial role in
determining the precise mass spectrum of the Higgses. In ${\cal N}=1$ supersymmetric theories in general, the K\"{a}hler potential
is renormalized and it is not possible to determine its precise form, especially in a strongly coupled setting. Hence, we will consider particular examples of K\"{a}hler potentials consistent with our general setup. Although the detailed results for the mass spectra and the resulting phenomenology will vary with the choice of the K\"{a}hler potential, we expect these examples to be representative of a large class of K\"{a}hler potentials, at least as far as qualitative features are concerned.

In order to motivate the choice of K\"{a}hler potential which we use later, consider the special case where the dimension of the Higgs field is
one. Then we can use the one-loop Coleman-Weinberg correction to the canonical K\"{a}hler potential
 \cite{Buchbinder:1994iw, Grisaru:1996ve}:%
\begin{equation}
K=K_{(0)}+K_{1-loop}=H^{\dag}H-\frac{1}{32\pi^{2}}\Tr\left(  \left\Vert
M\right\Vert ^{2}\log\frac{\left\Vert M\right\Vert ^{2}}{\Lambda_{UV}^{2}%
}\right)
\end{equation}
to capture the effects of
couplings to a hidden sector which gets its mass from Higgs vevs.
Here, the trace runs over the supermultiplets of the theory, $\left\Vert
M\right\Vert ^{2}$ are the masses squared of the various states, and
$\Lambda_{UV}$ is a UV cutoff. In the weakly coupled limit,
$\left\Vert M\right\Vert ^{2} = 32 \pi^{2} \times \widehat{\delta} \times H^{\dag}H$
for some constant $\widehat{\delta}$. We can then write:
\begin{equation}
K_{\log}=H^{\dag}H\left(  1-\widehat{\delta}\log\frac{H^{\dag}H}{\Lambda_{(0)}^{2}%
}\right)
\label{Klog}
\end{equation}
where $\Lambda_{(0)}$ is some high-scale where new physics come in.
The $K_{\log}$ K\"{a}hler potential can be viewed as a limiting case of a
K\"{a}hler potential:%
\begin{equation}
K_{\Delta}\equiv\left(  H^{\dag}H\right)  ^{1/\Delta}
\label{confkah}
\end{equation}
where $H$ has dimension $\Delta=1+\delta$, with the identification $\delta \sim \widehat{\delta}$.
Note that $K_{\Delta}$ provides a ``completion" of the metric in the sense that all higher powers in $K_{log}$ are resummed. Hence, important features of having a non-trivial scaling dimension may be missed in the leading log approximation. This will become clear in section \ref{weak}.

In writing the K\"{a}hler potential as $K_{\Delta}$ we have assumed that the dimension of $H^{\dag}\,H$ equals $2\Delta$. This is not true in general, but
provided it is close to $2 \Delta$, which is a conservative assumption when $\Delta$ is near one, our main conclusions will not change. For simplicity
of presentation, however, in what follows we shall take $H^{\dag}\,H$ to have dimension exactly $2 \Delta$.
In the small $\delta$ limit,  $x^{1/\Delta}=x(1-\delta\log x)$, which reproduces the behavior of the
weakly coupled example. Of course, in reality, the form of the K\"{a}hler
potential could be more complicated. The main qualitative condition we shall
be concerned with here is situations where the K\"ahler potential is
singular in the $H\rightarrow0$ limit and (\ref{confkah}) is a simple example satisfying this criterion.
The absence of a mass scale in equation (\ref{confkah}) also indicates that this is perhaps indicative of a conformal sector.
In realistic conformal theories, one does not expect the K\"ahler potential to
take exactly the form of $K_{\Delta}$. However, it is worth noting that
in some situations with $\mathcal{N}=2$ supersymmetry, this expression
is exact (see for example \cite{Gaiotto:2008nz} and \cite{HyperKahlerPOT}).

Self-consistency of our approximation dictates that the Higgs vev is
the dominant source of mass for at least some of the states of the hidden
sector. We note that even if the mass spectrum in the quasi-hidden sector is
not exactly supersymmetric, this behavior will continue to hold. For example,
while the MSSM\ scalars can have mass contributions
not proportional to the Higgs vevs, the fermion masses are proportional to the
Higgs vev. In the small $\delta$ limit, we can estimate these masses as
follows. Comparing the form of the Coleman-Weinberg potential to the form of
the K\"{a}hler potential $K_{\log}$, we identify:\footnote{Here we have
made an implicit assumption that the coupling associated to $H  \cO$ is an order one
number, with some natural notion of canonical normalization for the Higgs fields.}
\begin{equation}
M_{extra}\simeq4\pi v\times\sqrt{\delta}%
\end{equation}
where $v\simeq 246$ GeV is the Higgs vev.

What is the regime of validity of our setup? We expect that
our discussion is valid up to energy scales where one loop corrections can
destabilize the form of the effective potential. Indeed, since the potential arising from the deformation (\ref{deformation})
contains fractional powers of the fields, expanding around a fixed vev
will lead to a field theory with an arbitrary number of higher point
interactions. The higher-point interactions can renormalize the lower-point interactions by closing some of the external legs.
In order for these loop effects to not overwhelm the tree-level effects, the loop momenta must be bounded from above.
Hence, roughly viewing the mass scale for the Higgs vev $v$ as a decay
constant, we expect the effective potential to be valid up to a scale:%
\begin{equation}
\Lambda_{(0)}\simeq4\pi v\simeq3\text{ TeV.}%
\end{equation}
Let us note that this is a conservative assumption, the range
of validity may be higher due to cancellations from superpartners.

\subsection{The Extra States\label{ssec:EXTRAS}}

As we have seen, in the DSSM, there are extra states which get their mass from
the couplings:%
\begin{equation}
\int d^{2}\theta\text{ }\left (H_{u} \cO_{u}+H_{d} \cO_{d} \right ).
\end{equation}
Let us discuss some further properties of these extra states. We
assume that some of the hidden sector states get most of their mass
from the Higgs vevs. One could imagine that supersymmetry breaking causes some of the
bosons to be heavier than their fermionic superpartners, just as in the MSSM.
This will not change the fact that some states get their
mass mainly from the Higgs vevs.

Depending on the precise size of these mixing terms, the mass of these extra
fermionic states can be either above or below the TeV scale. In principle, one can
introduce additional mass terms for these vector-like states in order to remove them from the low-energy spectrum. This is possible but runs counter to our goal of having light states charged under $SU(2) \times U(1)$ with mass induced primarily by the Higgs vev\footnote{As explained in the Introduction, this
can naturally happen in some string constructions.}.

Of course, we do not know the precise form of the hidden sector states. However, in the weakly coupled limit where a Coleman-Weinberg like analysis is valid, there is a qualitative similarity to a
vector-like fourth generation with field content $Q^{(4)}\oplus\widetilde
{Q}^{(4)}$, $U^{(4)}\oplus\widetilde{U}^{(4)}$, $D^{(4)}\oplus\widetilde
{D}^{(4)}$, $L^{(4)}\oplus\widetilde{L}^{(4)}$, $E^{(4)}\oplus\widetilde
{E}^{(4)}$, where the superscript $(4)$ indicates only that these are all fourth-generation states. Here, the operators $\cO_{u}$ and $\cO_{d}$ are:%
\begin{align}
\cO_{u}  &  =\kappa_{QU}Q^{(4)}U^{(4)}+\widetilde{\kappa}_{QD}\widetilde
{Q}^{(4)}\widetilde{D}^{(4)}+\widetilde{\kappa}_{LE}\widetilde{L}%
^{(4)}\widetilde{E}^{(4)}+\kappa_{LS}L^{(4)}S_{u} + ...\label{Oumix}\\
\cO_{d}  &  =\widetilde{\kappa}_{QU}\widetilde{Q}^{(4)}\widetilde{U}%
^{(4)}+\kappa_{QD}Q^{(4)}D^{(4)}+\kappa_{LE}L^{(4)}E^{(4)}+\widetilde{\kappa
}_{LS}\widetilde{L}^{(4)}S_{d} + ...
\end{align}\label{4thgen}
We have included an additional vector-like pair of hidden sector singlets
$S_{u}\oplus S_{d}$. The values of the $\kappa$'s affect
the masses of these additional states. We see that all these fields
can be expected to participate in multiple interactions.

It is important to note that the description of the ${\cal O}'s$ above as a vector-like fourth generation is at best
very rough and \emph{qualitative}. The \textquotedblleft$...$\textquotedblright\ reflects
our lack of knowledge of the structure of ${\cal O}_u$ and ${\cal O}_d$.
For example, the beta function contributions to the SM gauge couplings
coming from the hidden sector are in general non-integral (as in
\cite{HVW}), hence the hidden sector states should not literally be viewed as a vector-like fourth
generation.

The presence of additional states charged under the SM\ gauge group will in
turn affect the running of the gauge couplings. One concern is that if these
extra states do not fill out GUT multiplets, gauge coupling unification will
be distorted. Another concern is that even if these states do fill out complete
GUT multiplets, the presence of many additional states can produce Landau
poles at low energies.

The first concern is naturally solved in models where $SU(5)_{GUT}$ is a
flavor symmetry. The second concern can also be bypassed by
taking $\delta$ sufficiently small, since $\delta$ is what
enters into the numerator of the NSVZ beta function. As found in \cite{HVW},
there exist explicit string theory scenarios where gauge coupling unification can be
retained, even when the threshold is on the order of the weak scale (see also
\cite{Donkin:2010ta}).\footnote{Let us note that in the models in
\cite{HVW}, additional couplings between the Standard Model and quasi-hidden
sector were also included, which in turn further shift the scaling dimensions
of the MSSM\ fields. A small variant on the models of \cite{HVW} can be
arranged by switching off these additional F-term mixing terms. Alternatively, these additional couplings can be retained
and their effect examined. In this work, we omit these terms for simplicity. The main point here is
that the requirements for gauge coupling unification can be met in
explicit models.}

Finally, the hidden sector may also include additional states which are not
charged under the SM\ gauge group. This is the situation, for example, in the
probe D3-brane theories considered in \cite{HVW}. Because not all such states couple to the SM, one
expects that some of these states will be lighter than their counterparts
charged under $G_{SM}$. We shall return to the
phenomenological consequences of these additional states in section
\ref{sec:collider}.

\subsection{The Range of $\delta$ \label{ssec:RANGE}}

In this section we discuss bounds on the Higgs dimensions. We phrase our discussion in terms of
$\Delta_{u},\Delta_{d}$ and the excess dimension $\delta = (\Delta_u + \Delta_d)/2 - 1$.
There is a general lower bound coming from the fact that the extra charged
states cannot be too light. Taking for concreteness $M_{top}\simeq175$ GeV $\lesssim
M_{extra}\simeq\Lambda_{(0)}\times\sqrt{\delta}\simeq(3$ TeV$)\times
\sqrt{\delta}$, it is clear that $\delta$ is at least of order $3\times10^{-3}$.
Assuming the couplings $H_{u}QU$ and $H_{d}QD$ are generated at the GUT scale,
making the Higgs a higher dimension operator means that the top and
bottom Yukawas will experience some conformal suppression:%
\begin{equation}
\lambda_{top}\left(  M_{top}\right)  \sim3\times\left(  \frac{M_{H}}{M_{GUT}%
}\right)  ^{\Delta_{u}+\Delta_{Q}+\Delta_{U}-3}\text{, }\lambda_{bot}\left(
M_{bot}\right)  \sim3\times\left(  \frac{M_{H}}{M_{GUT}}\right)  ^{\Delta
_{d}+\Delta_{Q}+\Delta_{D}-3}.%
\end{equation}
The factor of three comes from running from the GUT scale
down to the weak scale. This assumes that flavor is generated
at the GUT\ scale, as is natural in F-theory GUT\ models. For example, taking
$M_{H}/M_{GUT}\sim10^{-13}$, we see that $3\times\left(  M_{H}/M_{GUT}\right)
^{0.01}\sim2$, while $3\times\left(  M_{H}/M_{GUT}\right)  ^{0.1}\sim0.2$. Thus, $0.01 \lesssim \delta \lesssim 0.1$
is naturally consistent with a large top Yukawa. A relevant issue in this regard is whether other SM fields are
assumed to mix slightly with the D3-brane states.
This is true in the examples considered in \cite{HVW}, although not strictly necessary for a field theory
realization of the small $\delta$ regime. In this
work, for simplicity and concreteness, we consider UV deformations where only the Higgs fields mix
with the hidden sector, since otherwise one  has to also include non-oblique corrections to electroweak precision observables. However, it is worth noting that such mixings can be potentially beneficial. For
example, when the $\overline{5}_{M}$ also has a non-trivial scaling dimension,
this can lead to conformal suppression of the bottom Yukawa relative to the
top Yukawa, so that a low $\tan\beta$ can be easily arranged.

In summary, we see that the viable range of values for$~\delta$ is:%
\begin{equation}
0.01\lesssim\delta\lesssim0.1.
\end{equation}
As we explain later, in the above region of $\delta$, the DSSM can be consistent
with precision electroweak constraints without much tuning. Quite fortuitously, this is also the range which naturally occurs in the models of
\cite{HVW}!

\section{Mass Scales in the DSSM} \label{sec:SPECTRUM}

In this section we estimate the Higgs and gauge boson masses in the
DSSM. Based on the general considerations of section \ref{sec:SETUP}, we focus
primarily on the small $\delta$ regime. Since we do not know the
precise form of the K\"ahler potential in general, it is more reliable to
provide general scaling estimates. In subsection \ref{ssec:MEXICAN} we show
that the Higgs vev can naturally be far lower than the lightest Higgs mass.
We then study the gauge boson masses in the presence of a non-trivial
K\"ahler potential. See Appendix B for some explicit examples.

\subsection{Symmetry Breaking in a Squeezed Mexican Hat \label{ssec:MEXICAN}}

We now study symmetry breaking for the DSSM Higgs potential. The basic
idea is that there are four relevant operators in this potential: two soft mass terms, the $B\mu$-term, and the
supersymmetry-preserving $\mu$-term. Although there is also a D-term potential, it is subdominant because
it comes from an irrelevant operator. In the MSSM, the very irrelevance of the D-term is responsible for the
Little Hierarchy problem. By contrast, in the DSSM, the Higgs mass can be far higher than the mass scale set by its vev.

\begin{figure}[t!]
\begin{center}
\includegraphics[
height=2.5564in,
width=3.9885in
]{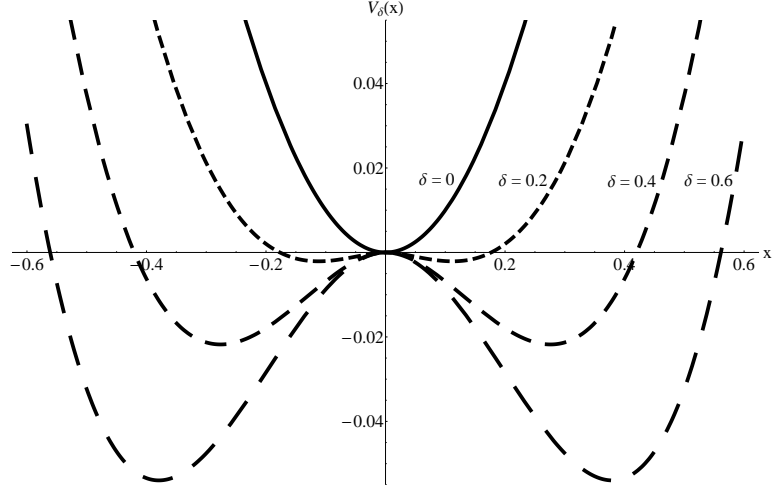}
\end{center}
\caption{Cartoon plot of the family of functions $V_{\delta}(x) = - \vert x \vert^{2} + 2 \vert x \vert^{2 + 2 \delta}$ for various choices of $\delta$. At $\delta$ exactly zero, the non-zero minimum disappears. For $\delta \neq 0$, this leads to a squeezed minimum close to the origin of field space.}%
\label{mexicans}%
\end{figure}

Before discussing a particular example, let us first explain in general terms
why the form of the effective potential changes so drastically
compared to the standard MSSM Higgs potential. The cause of
this change can be traced to a competition between terms of degree
close to two. At small field range, these terms dominate over the
irrelevant D-term contribution. For a field $H$ of small excess
dimension $\delta = \Delta - 1$, the role of the quartic in the standard Mexican
hat potential is, at very small field range, instead taken up by the $\mu$-term:
\begin{equation}
V_{eff} \supset \vert \mu \vert^{2} \vert H \vert^{2 + 2 \delta}.
\end{equation}
The role of the tachyonic mass term in the usual Mexican hat potential can in principle be played by lower degree terms such as $\vert H \vert ^{2 - 2 \delta}$ or $\vert H \vert^{2}$, which respectively originate from the soft masses or the $B \mu$-term. Only one such term is really necessary to produce a stabilized potential, but in a generic situation we can expect both to be present. These tachyonic contributions are at least as relevant as a usual mass term in a Mexican hat potential. Balancing the $\mu$-term against one of these contributions, the net effect is a squeezed Mexican hat potential. See figure \ref{mexicans} for the general behavior of such functions as $\delta$ is varied.

Having illustrated the general idea, let us now consider a toy model which exhibits this behavior.
Consider a single chiral superfield $X$ of dimension $\Delta_{X}$ and K\"{a}hler potential $K=(X^{\dag
}X)^{1/\Delta_X}$. In what follows we assume that the dimension of $X^{\dag}X$ is $2\Delta_X$, which is a good approximation
in many examples but which may be relaxed, see \cite{Rattazzi:2008pe, Poland:2010wg,
Fitzpatrick:2011hh}. In this setting, we now show that the analogue of
the soft supersymmetry breaking terms in the Higgs sector lead to a mass for
$X$ which is far above the vev $\vert \langle X \rangle \vert ^{1/\Delta_{X}}$.
With this choice of K\"{a}hler potential, the K\"ahler metric is
\begin{equation}
g_{X \overline X} = \frac{1}{\Delta_X^2}\frac{|X|^{2/\Delta_X}}{|X|^2}
\end{equation}
and the effective kinetic term is:%
\begin{equation}
L_{kinetic}= - \frac{1}{\Delta_{X}^{2}}\frac{\left\vert X\right\vert ^{2/\Delta_{X}}%
}{\left\vert X\right\vert ^{2}}\left\vert \partial_{\mu}X\right\vert ^{2}.%
\end{equation}
Motivated by the arguments in section \ref{sec:SETUP} we study the following scalar potential:%
\begin{align}
V  &  =m^{2}g_{X\overline{X}}\left\vert X\right\vert ^{2}-B\left\vert
X\right\vert ^{2}+\left\vert \mu\right\vert ^{2}g^{X\overline{X}}\left\vert
X\right\vert ^{2} \nonumber \\
&  =\frac{m^{2}}{\Delta_{X}^{2}}\left\vert X\right\vert ^{2/\Delta_{X}%
}-B\left\vert X\right\vert ^{2}+\Delta_{X}^{2}\left\vert \mu\right\vert
^{2}\left\vert X\right\vert ^{(4\Delta_{X}-2)/\Delta_{X}}\nonumber \\
&  \simeq A\left\vert X\right\vert ^{2-2\delta_{X}}-B\left\vert X\right\vert
^{2}+C\left\vert X\right\vert ^{2+2\delta_{X}}  \label{ABCpot}%
\end{align}
where $A \equiv m^2/\Delta_X^2 \simeq m_0^2\,\Lambda_{soft}^2\,(1-2\delta_X)$, $B \simeq B_0\,\Lambda_{soft}^{2-2\delta_X}$,
and  $C \equiv |\mu|^2 \Delta_X^2 \simeq |\mu_0|^2\,\Lambda_{soft}^{2-4\delta_X}\,(1+2\delta_X)$. Here we have introduced dimensionless order one constants
$m_0, B_0$ and $\mu_0$ as per our discussion in section \ref{sec:SETUP}. In comparison to a standard
Mexican hat potential, we observe that the exponents are compressed.
This leads to an interesting change in the behavior of the potential. The critical
points of this potential satisfy:%
\begin{equation}\label{critical}
\left\vert X_{\ast}^{(1)}\right\vert ^{2(1-\delta_{X})}=0;\;\left\vert X_{\ast}^{(2)}\right\vert ^{2(1-\delta_{X})}=\Lambda_{soft}^{2}\times\left(
\tilde{q}_{0}\right)  ^{1/\delta_X};\;\left\vert X_{\ast}^{(3)}\right\vert ^{2(1-\delta_{X})}=\Lambda_{soft}^{2}\times\left(
q_{0}\right)  ^{1/\delta_X}.
\end{equation}
where $q_{0}$ and $\tilde{q}_0$, to leading order in $\delta_{X}$, are given by :%
\begin{equation}
q_{0}\simeq \frac{B_0+t_0}{2|\mu_0|^2},\;
\tilde{q}_{0}\simeq \frac{B_0-t_0}{2|\mu_0|^2};\;t_0\equiv \sqrt{B_0^{2}-4\,m_0^2\,|\mu_0|^2}%
\end{equation}
The critical point at zero is special since the second derivative diverges
there, reflecting the singularity of the K\"ahler metric at the origin
of field space. Provided the parameters $q_0$ and $\tilde{q}_{0}$ are both
positive, the other two critical points give rise to a maximum and minimum at $\left\vert X_{\ast}^{(2)}\right\vert $ and $\left\vert X_{\ast}^{(3)}\right\vert$, respectively. Let us note that a symmetry-breaking vev can be achieved even when $m = 0$, or alternatively, when $B = 0$ and $m^{2} < 0$. In these limits, the ``dip'' at the origin of field space again flattens out.

The crucial point to note is that the vev is exponentially suppressed relative to
$\Lambda_{soft}$ when $q_{0}<1$.\footnote{Note that we want $q_{0}$ to not be \emph{much} smaller than one, otherwise there will be a huge hierarchy between the vev and $\Lambda_{soft}$. The phenomenologically relevant mass scales are discussed in section \ref{sec:MASSSUMMARY}.} In this case, the mass squared of the canonically normalized field $X$ can be computed to leading order in $\delta_{X}$:
\begin{equation}
M_{X}^{2} \simeq \Lambda_{soft}^{2}%
\times\delta_{X}\times t_{0}.
\end{equation}
Note that the mass is only suppressed relative to $\Lambda_{soft}$ by a factor $\sqrt{\delta_X}$, hence the vev is exponentially
suppressed relative to the mass. The proximity of low-degree
terms explains why the contribution from an ordinary quartic term is
negligible; in the low energy theory it is far less relevant than the lower
degree terms. What happens when $q_{0} > 1$? In that case, our description has broken down since the purported
vev is far above $\Lambda_{soft}$.

The above analysis can be straightforwardly generalized to models with two Higgs doublets. Symmetry
breaking with the Higgs fields is encapsulated in terms of the vevs
$v_{u}=\left\langle H_{u}^{(0)}\right\rangle $ and $v_{d}=\left\langle
H_{d}^{(0)}\right\rangle $, which are related to the two mass scales
$v_{u}^{1/\Delta_{u}}$ and $v_{d}^{1/\Delta_{d}}$. It is helpful to
parameterize these scales in terms of the characteristic mass scale $v$ and a
dimensionless ratio:%
\begin{equation}
\frac{v}{\sqrt{2}} = \sqrt{v_{u}^{2/\Delta_{u}}+v_{d}^{2/\Delta_{d}}}\text{, }\tan\beta
=\frac{v_{u}^{1/\Delta_{u}}}{v_{d}^{1/\Delta_{d}}}\text{.}%
\end{equation}
The general form of the effective potential as a function of $v$ can then be fixed
by scaling arguments to be:%
\begin{equation}
V\left(  v,\tan\beta\right)  =m^{2}v^{2} - Bv^{2\Delta}+\left\vert
\mu\right\vert ^{2}v^{4 \Delta - 2}%
\end{equation}
where the dependence on $\tan\,\beta$ is absorbed in the coefficients. We are implicitly assuming that $\tan\,\beta$ is a free parameter, the changes in which can be compensated by changes in the supersymmetry breaking parameters\footnote{The same is done in phenomenological studies of the MSSM.}. The astute reader may notice that the exponents of $v$ are slightly different from those discussed earlier in the context of the toy model involving $X$. This is because $v$ has canonical dimensions of mass, whereas $X$ did not. Here, the scaling dimensions are $m^{2}%
\sim\Lambda_{soft}^{2}$, $B\sim\Lambda_{soft}^{4-\Delta_{u}-\Delta_{d}}$,
$\mu\sim\Lambda_{soft}^{3-\Delta_{u}-\Delta_{d}}$, and $\Delta = (\Delta_{u}+\Delta_{d}) / 2 $.
Following the same discussion as before, we see that the characteristic vev
and mass squared of the lightest Higgs $H$ is:
\begin{equation}
v^{2}\simeq\Lambda_{soft}^{2}\times\left(  q_{0}\right)  ^{1/\delta}\text{,
}M_{H}^{2}\simeq\Lambda_{soft}^{2}\times\delta .
\end{equation}

In contrast to the usual MSSM, not all of the soft mass terms or even the quartic D-terms are necessary to generate a symmetry-breaking minimum. The
$\mu$-term induces a term of degree $2 + 4 \delta$, and plays the role of a large quartic term. The soft masses and $B \mu$-term serve instead as
tachyonic masses. The $B\mu$-term always induces a saddle point, generating a tendency to roll away from the origin of field space. Hence, one can in principle do without the soft mass terms. Conversely, one can consider scenarios where the $B \mu$-term is absent, and the soft mass terms are tachyonic. These two possibilities are especially tractable and are treated in more detail in Appendix B.

Summarizing, we see that the mass of the lightest Higgs is slightly suppressed
relative to $\Lambda_{soft}$, while its vev is exponentially suppressed. The
vev determines the mass of the gauge bosons, leading to a Higgs mass which can
naturally be above the gauge boson masses. Hence, we obtain the general
relation:%
\begin{equation}
\frac{M_{H}}{M_{V}}\simeq\frac{\sqrt{\delta}}{g}\times\left(  \sqrt{q_{0}%
}\right)  ^{-1/\delta}.%
\end{equation}

\subsubsection{The Weakly Coupled Limit}\label{weak}

Before moving on to discussing gauge boson masses, it is worth understanding the weakly coupled limit in which the Higgs fields have dimension exactly one.
In this case, it is tempting to use the Coleman-Weinberg K\"{a}hler potential $K_{log}$ in (\ref{Klog}) to determine the effective potential and the
corresponding vacuum. This is certainly possible to do; however this leads to the following solution for the vev:
\begin{equation}\label{weaklimit}
\delta\,\log\left(\frac{v^2}{\Lambda_{(0)}^2}\right) \simeq \left(\frac{B - |\mu|^2 - m^2}{|\mu|^2 - m^2}\right) = {\cal O}(1),
\end{equation}
The parameters appearing in (\ref{weaklimit}) have the usual dimensions since we are in the weakly coupled limit. Since the RHS is generically ${\cal O}(1)$ or greater as it is independent of $\delta$, this implies that higher order corrections to the K\"{a}hler potential must be taken into account\footnote{Note that precisely the same argument was given by Coleman and Weinberg for not trusting the symmetry breaking solution in one of the examples in \cite{Coleman:1973jx}.}.  Thus, the full K\"{a}hler potential $K_{\Delta}$, which provides a UV completion of $K_{log}$, is needed to study the vacuum of the theory.

\subsection{Gauge Boson Masses\label{ssec:SYMMBREAK}}

Let us now discuss some further features of the gauge bosson mass spectrum.
From the perspective of the hidden sector, the Standard Model appears as a
weakly gauged flavor symmetry. Since the Higgs fields are, by definition,
charged under $SU(2)_{L}\times U(1)_{Y}$, we can write down the leading order
coupling between the Higgs modes and gauge bosons, as in a general
$\mathcal{N}=1$ supersymmetric field theory:%
\begin{equation}
T_{Higgs}= - g_{i\overline{j}} D_{\mu}\Phi^{i} D^{\mu}\Phi^{\dag \overline{j}}.%
\end{equation}
Here, $\Phi^{i}$ is shorthand for the vector of chiral
superfields $\Phi=(H_{u}^{(+)},H_{u}^{(0)},H_{d}^{(0)},H_{d}^{(-)})$,
$g_{i\overline{j}}$ is the K\"{a}hler metric. Since the flavor symmetry is assumed to be
only weakly gauged, the covariant derivative for this four-component
vector is given by the usual covariant derivative for the two Higgs doublets $H_u$ and $H_d$.

A remarkable ``accident'' of the Standard Model and MSSM is that the heavy gauge bosons $(W^{(+)},Z^{(0)},W^{(-)})$
form an approximate triplet under a custodial $SU(2)$ symmetry. This is no
longer guaranteed when the K\"ahler potential is of a general form. We refer to the ``classical contribution''
from expanding the K\"ahler potential around a fixed vev as $M_{W,cl}$ and $M_{Z,cl}$. It is also
convenient to introduce the ratio:
\begin{equation}
\rho_{cl}=\frac{M_{W,cl}^{2}}{c_{W}^{2}M_{Z,cl}^{2}} \label{weinbergangles}%
\end{equation}
where $c_{W}^{2}=\frac{g^{2}}{g^{2}+g^{\prime2}}$ and $g$ and $g^{\prime}$ are the gauge couplings of $SU(2)_{L}$
and $U(1)_{Y}$, respectively.\footnote{Here we neglect convention-dependent subtleties connected with the renormalization of electroweak observables, and in particular, which quantities we treat as fixed inputs, and which are taken as scale
dependent (see e.g. \cite{Peskin:1990zt} and the review \cite{Wells:2005vk} for discussion on this point).}
Violation of custodial $SU(2)$ is controlled by the order parameters $\delta_{u}=\Delta_{u}-1$ and $\delta_{d}=\Delta_{d}-1$.
Let us note that the actual $\rho$-parameter of the DSSM also includes radiative contributions, which are not included in this ``classical expression''.
We return to this in section \ref{sec:CONSTRAINTS}.

Consider first the case where the Higgs fields both have dimension $\Delta$ and the
K\"{a}hler potential is $K_{\Delta}=(H_{u}^{\dag}H_{u}%
+H_{d}^{\dag}H_{d})^{1/\Delta}$. Taking the Higgs vevs to be $v_{u}$ and $v_{d}$, the
resulting mass terms for the gauge bosons are:
\begin{equation}
M_{W,cl}^{2}=\frac{g^{2}}{2\Delta}\times\left(  v_{u}^{2}+v_{d}^{2}\right)
^{1/\Delta}\text{, }M_{Z,cl}^{2}=\frac{M_{W,cl}^{2}}{c_{W}^{2}}\times
\frac{\left(  \left(  v_{u}^{2}+v_{d}^{2}\right)  ^{2}+4v_{u}^{2}v_{d}%
^{2}(\Delta-1)\right)  }{\Delta\left(  v_{u}^{2}+v_{d}^{2}\right)  ^{2}}
\label{gaugeboson}%
\end{equation}
so that $\rho_{cl}$ is:%
\begin{equation}
\rho_{cl}\simeq1+\left(  1-\frac{4v_{u}^{2}v_{d}^{2}}{\left(  v_{u}^{2}%
+v_{d}^{2}\right)  ^{2}}\right)  \delta
\end{equation}
and we have expanded in small $\delta$. In the limit
of low $\tan\beta$ and large $\tan\beta$, we have:%
\begin{align}
\tan\beta &  \rightarrow1:\rho_{cl}=1+\delta\times\varepsilon^{2}%
\label{tandisc}\\
\tan\beta &  \rightarrow\infty:\rho_{cl}=1+\delta.
\end{align}
The reason for the extra suppression by $\varepsilon$ in the first case is
that $K_{\Delta}$ preserves an $SU(2)_{L}\times SU(2)_{R}$ under which the
Higgs fields transform in the $(2,2)$. When $\tan\beta\rightarrow1$, a
diagonal $SU(2)$ is preserved, so the gauge bosons fill out a triplet of
the approximate custodial $SU(2)$.

As another example, consider a perhaps more generic situation where $H_u$
and $H_d$ have respective dimensions $\Delta_u$ and $\Delta_d$,
with K\"{a}hler potential:
\begin{equation}
K=(H_{u}^{\dag}H_{u})^{1/\Delta_{u}}+(H_{d}^{\dag}H_{d})^{1/\Delta_{d}}.
\end{equation}
Computing the gauge boson masses in this case yields, as a function of the
Higgs vevs:
\begin{equation}
M_{W,cl}^{2}=\frac{g^{2}}{2}\times\frac{\Delta_{d} v_u^{2 / \Delta_u} + \Delta_{u} v_d^{2 / \Delta_d}
}{\Delta_{u}\Delta_{d}} \text{, }M_{Z,cl}^{2}=\frac{g^{2} + g^{\prime 2}}{2}\times\frac{\Delta_{d}^{2} v_{u}^{2 / \Delta_u} +\Delta_{u}^{2} v_d^{2 / \Delta_{d}}}{\Delta_{u}^{2}\Delta_{d}^{2}}.
\end{equation}
The value of $\rho_{cl}$ is now:
\begin{equation}
\rho_{cl}\simeq1+\frac{\tan^{2}\beta\times\delta_{u}+\delta_{d}}{\tan^{2}%
\beta+1}.
\end{equation}
Observe that for this choice of K\"{a}hler potential, the more generic
estimate is again $\rho_{cl}\simeq1+c\times\delta$, where $c$ is a number slightly less than one, and $\delta$ is the characteristic dimension of
the two Higgs fields. The two limiting cases are:
\begin{align}
\tan\beta &  \rightarrow1:\rho_{cl}=1+\frac{\delta_{u}+\delta_{d}}{2}\\
\tan\beta &  \rightarrow\infty:\rho_{cl}=1+\delta_{u}.
\end{align}

\subsection{Summary of Mass Scales}\label{sec:MASSSUMMARY}

To orient the reader, let us now summarize the various mass scales of the DSSM
in the small $\delta$ regime. The estimates of these scales will appear in our estimates of the
behavior of the physical potential. We first express all quantities in terms
of the single scale $\Lambda_{soft}$.

Proceeding up in energy scales, we first encounter the gauge boson masses. The
mass of the $W$-boson is:
\begin{equation}
M_{W} = \frac{g\,v}{2}\simeq \frac{g}{2} \times \Lambda_{soft} \times \left(  \sqrt{q_{0}}\right)
^{1/\delta }%
\end{equation}
where $q_{0}<1$ is an order one number. We see
that $M_{W}$ is exponentially suppressed relative to $\Lambda_{soft}$. The
lightest Higgs $H$ and the extra states are both heavier than the gauge boson mass, but
suppressed relative to $\Lambda_{soft}$ and $\Lambda_{(0)}$  by a factor of $\sqrt{\delta}$:%
\begin{align}
M_{H}  &  \simeq\Lambda_{soft}\times\sqrt{\delta}\\
M_{extra}  &  \simeq4\pi v\times\sqrt{\delta} = \Lambda_{(0)}\times \sqrt{\delta}.
\end{align}
Insofar as we expect $\Lambda_{soft}$ to be less than
$\Lambda_{(0)}$, we see that the Higgs will be somewhat lighter than the extra
states for comparable Yukawas. Note, however, that this separation is independent of $\delta$. We
also see that the mass of the Higgs relative to the gauge bosons is:%
\begin{equation}
\frac{M_{H}}{M_{W}}\simeq\frac{2}{g}\times\sqrt{\delta}\times\left(
\sqrt{q_{0}}\right)  ^{-1/\delta}.
\end{equation}
In other words, we see that with $q_0 < 1$, as $\delta$ decreases the Higgs becomes heavier
than the $W$-boson. However, since the ratio is quite sensitive to the value of $\delta$, the mass of the Higgs can
either be close to the LEP bound of $114$ GeV, or far heavier.

It is also interesting to study the behavior of $\Lambda_{soft}$ viewed as a
function of the fixed mass scale $M_{W}$ and the parameters $g$, $q_{0}$ and
$\Delta$:%
\begin{equation}
\Lambda_{soft}\simeq\frac{2\,M_{W}}{g}\times\left(  \sqrt{q_{0}}\right)^{-1/\delta}.
\end{equation}
Note that since $\Lambda_{soft}\leq\Lambda_{(0)}\simeq3$ TeV, naively it would seem that the Higgs mass can be increased by an arbitrary amount by
increasing $\delta$. However, the above formulae rely on a classical analysis of the Higgs potential where the SM gauge group is assumed to be weakly gauged. As a conservative estimate, for Higgs masses above around a TeV, weak gauge boson scattering ceases to be perturbative and partial-wave unitarity is violated \cite{Lee:1977eg}. Hence, when the Higgs mass becomes close to a TeV, the framework described in this work ceases to be valid. This breakdown in the analysis occurs when $\Lambda_{soft} \sim \Lambda_{0} \sim 4 \pi v$. Stated differently,
the approximation scheme remains self-consistent provided:
\begin{equation}
4 \pi > \left(  \sqrt{q_{0}}\right)^{-1/\delta}.
\end{equation}
For $\delta = 0.01$, this implies $q_0 \gtrsim 0.95$, while for $\delta = 0.1$, this implies $q_{0} \gtrsim 0.6$.
Thus, $q_0$ is required to be close to one and less than one for consistency. We do not view this as a
fine-tuning, since we are simply solving for $q_{0}$ as a function of $\delta$. Indeed, when this condition is not satisfied, it simply means
perturbation theory has broken down.

\section{Precision Constraints \label{sec:CONSTRAINTS}}

In the previous section we observed that the presence of a singular K\"{a}hler
potential naturally increases the mass of the lightest Higgs boson relative to
the $W$- and $Z$-boson masses. It is well known that in the absence of any
additional physics, the precision electroweak fit favors a light $\sim 100$ GeV elementary Higgs.

In this section we study the consequences of such constraints in the DSSM. We
mainly focus on the contributions to the oblique parameters $S$ and $T$
\cite{Peskin:1990zt} as the $U$ parameter is expected to be much smaller.
Present electroweak fits yield an ellipse in the $S-T$ plane \cite{PDG} which
roughly lies along a $45^{\circ}$ angle with major axis going from the lower
left to the upper right. Increasing the mass of the Higgs leads to a
contribution to $S$ and $T$ which moves these values down and to the right of
the ellipse. Hence, a positive contribution to $T_{UV}$ and a negligible
positive or negative contribution to $S_{UV}$ is necessary in order to pass
back within the $68\%$ confidence level ellipse. To frame our discussion,
recall that to lowest order $S$ is defined as the coefficient of the operator
$\frac{g\,g^{\prime}}{16\pi}\,\Pi^{\dag}W_{\mu\nu}^{3}\,\Pi\,B^{\mu\nu}$,
while $T$ is the coefficient of $\frac{e^{2}}{8\pi\,c_{W}^{2}}\,v^{2}%
\left\vert \Tr_{SU(2)}(\tau^{(3)}\,\Pi\,D_{\mu}\Pi^{\dag})\right\vert ^{2}$,
where $\Pi=e^{i\frac{\vec{\pi}\cdot\vec{\tau}}{v}}$ and the $\tau^{(a)}$ are
spin $1/2$ generators of $SU(2)$. Our conventions are that we consider the
contributions to these operators from all effects beyond the Standard Model.
In our conventions, the Standard Model values are then $S_{SM}=0$, $T_{SM}=0$.

Let us preface our discussion with a general comment: Since we have only
imperfect knowledge of the hidden conformal sector, and since it is strongly coupled, we shall only aim for order of
magnitude estimates. It would be of interest to turn the discussion around,
and use constraints from electroweak precision as a way to obtain insight about the properties of
the hidden sector of the setup. The main conclusion we draw is that
the contributions from the minimal inputs necessary to define the DSSM can
accommodate constraints from precision electroweak. Also, we will first discuss the case where custodial $SU(2)$ is not preserved, and then discuss situations in which it could be approximately preserved.

We divide the contribution to $S$ and $T$ in two pieces - one is the \textquotedblleft classical\textquotedblright%
\ UV contribution to the electroweak observables coming from expanding around a
fixed Higgs vev, and the other is IR radiative contributions in the Higgs
sector at energies of the order of Higgs masses and below:
\begin{align}
S &  =S_{IR}+S_{UV}\\
T &  =T_{IR}+T_{UV}%
\end{align}
For the radiative corrections from the Higgs sector, the Higgs fields are treated as a collection of
ordinary weakly coupled fields. We expect the division into two pieces to at least roughly capture the physics.
This is because the classical analysis of the scalar potential differs from the standard weakly coupled models due to a non-trivial K\"{a}hler potential, giving rise to a  ``classical UV" contribution.  However, the fluctuations at low energies are close to that of weakly coupled
fields (since $\Delta$ is close to 1) giving rise to a ``radiative IR" contribution which is similar to that in weakly coupled models.

Consider first the IR contributions. These radiative contributions are of the
standard type which have been extensively studied for two Higgs doublet
models. We adopt the approximation scheme used in Appendix A of
\cite{Harnik:2003rs}, and closely follow the discussion provided there. We
split up the Higgs mass eigenstates $h^{0}$, $H^{0}$, $H^{\pm}$ and $A^{0}$,
into a Standard Model-like scalar Higgs $\widetilde{h}^{0}$ and an $SU(2)_{L}$
doublet:%
\begin{equation}
H=\left[
\begin{array}
[c]{c}%
H^{+}\\
\frac{1}{\sqrt{2}}\left(  \widetilde{H}^{0}+iA^{0}\right)
\end{array}
\right]
\end{equation}
where $\widetilde{H}^{0}=\cos\left(  \beta-\alpha\right)  h^{0}-\sin\left(
\beta-\alpha\right)  H^{0}$ and $\widetilde{h}^{0}$ is orthogonal to
$\widetilde{H}^{0}$. Our conventions for Higgs mixing angles are as in
\cite{MartinPrimer}. Since $\widetilde{h}^{0}$ is not a mass
eigenstate, it will have contributions from both $h^{0}$ and $H^{0}$ which can
be treated in terms of a weighted sum. Similarly, $\widetilde{H}^{0}$ is a
linear combination of mass eigenstates. The full contribution to $S_{IR}$ and
$T_{IR}$ can then be written, in this approximation, as a sum of the singlet
and the doublet contributions:%
\begin{equation}
S_{IR} \simeq S_{IR}^{\text{singlet}}+S_{IR}^{\text{doublet}}\,\, , \,\, T_{IR}  \simeq T_{IR}^{\text{singlet}}+T_{IR}^{\text{doublet}}
\end{equation}
where the contribution from the singlet can be approximated as (see e.g.
\cite{Peskin:1990zt}):
\begin{equation}
S_{IR}^{\text{singlet}} \simeq\frac{1}{6\pi}\log\frac{\overline{M}_{H}}{M_{Z}}\,\, , \,\, T_{IR}^{\text{singlet}} \simeq-\frac{3}{8\pi c_{W}^{2}}\log\frac
{\overline{M}_{H}}{M_{Z}}
\end{equation}
where $\overline{M}_{H}$ is the characteristic mass scale for the singlet
Higgs, which behaves roughly like a SM higgs. For example, using a representative value $\overline{M}_H = 500$ GeV (see Appendix B for a rough sense of the mass scales involved), we find :
\begin{equation} S_{IR}^{\text{singlet}} \approx +0.09;\;
T_{IR}^{\text{singlet}} \approx -0.26
\end{equation}
The contribution from the doublet is somewhat more involved. It is given by:
\begin{align}
S_{IR}^{\text{doublet}} &  \simeq\sin^{2}(\beta-\alpha)\,F(M_{H^{\pm}%
},M_{H^{0}},M_{A^{0}})+\cos^{2}(\beta-\alpha)\,F(M_{H^{\pm}},M_{h^{0}%
},M_{A^{0}})\\
T_{IR}^{\text{doublet}} &  \simeq\frac{1}{16\pi\,M_{W}^{2}\,s_{W}^{2}%
}\,\left[  \sin^{2}(\beta-\alpha)\,G(M_{H^{\pm}},M_{H^{0}},M_{A^{0}})+\cos
^{2}(\beta-\alpha)\,G(M_{H^{\pm}},M_{h^{0}},M_{A^{0}})\right]  \nonumber
\end{align}
Here, $F$ and $G$ are complicated functions of their arguments (see
Appendix A of \cite{Harnik:2003rs}). However, qualitatively these contributions have the rough form:
\begin{equation}
F\approx c_{F}\times\log\frac{\overline{M}_{H}}{M_{Z}}\text{, }G\approx
c_{G}\times\overline{M}_{H}^{2}\log\frac{\overline{M}_{H}}{M_{Z}}%
\end{equation}
where the $c$'s are order $0.1-1$ constants, and $\overline{M}_H$ is a characteristic mass scale in the arguments of $F$ and $G$. Note that $F$ is logarithmic in the Higgs masses, while $G$ contains a quadratic piece. Depending on the Higgs mass spectrum, $c_{F}$ and $c_{G}$ can either be positive or negative. Using the explicit forms of $F$ and $G$ and various choices for the Higgs masses which could arise in the DSSM, we find that $S_{IR}^{\text{doublet}}$ can vary between -0.1 and +0.1. $T_{IR}^{\text{doublet}}$ is far more sensitive to the Higgs masses, it can be either positive or negative with magnitude varying from around $0.1$ to around $10$. A ``large" contribution $|T| \sim 10$ arises if custodial $SU(2)$ is violated in the Higgs spectrum. Let us note that having both $S$ and $T$ negative occurs only in a small range of parameter space, and in such cases, the magnitude of the parameters are bounded. Related scans of two Higgs doublet parameter space have been performed in \cite{Haber:2010bw} and very recently in \cite{Funk:2011ad}. In \cite{Funk:2011ad}, it was also found that for some range of parameters a negative $S$ and $T$ could also be arranged. In our case, the difference with the perturbative two-Higgs doublet model arises due to additional ``UV" contributions. For example, for the $T$ parameter, the ``classical UV" contribution coming from a non-trivial K\"{a}hler potential also violates custodial $SU(2)$ in general\footnote{Of course, it is possible to consider cases where custodial SU(2) is approximately conserved, which is also discussed.}; hence  the full contribution can be within the allowed limits if they cancel each other. As will be seen below, the contribution to $T_{UV}$ is positive, so a reasonable fit will require $T_{IR}$ to be negative. This fixes some features of the Higgs masses but is still rather model-dependent, depending on the details of the Higgs spectrum.

Let us now turn to the various UV contributions. Consider first the contributions to the $T_{UV}$. In the approximation
scheme we have adopted, the contribution from extra states of the
conformal sector has already been included via the effective K\"{a}hler potential. This
generates what we referred to in the previous section as the \textquotedblleft
classical\textquotedblright\ $\rho_{cl}$. This is because the effective
K\"{a}hler potential for the Higgs fields encapsulates the effects of
wave-function renormalization for the Goldstone modes via the interactions
$H_{u} \cO_{u}+H_{d} \cO_{d}$. Hence, the UV contribution to $T$ is estimated to be:
\begin{equation}
T_{UV}=x\times\frac{\delta}{\alpha}+...\label{TUV}%
\end{equation}
where $x$ is an order one positive number (typically smaller than one), and we have
used the relation $\rho=1+\alpha\,T$ (see e.g. \cite{Peskin:1990zt}).
The \textquotedblleft$...$\textquotedblright reflects possible additional contributions from supersymmetry breaking
effects which, as in the MSSM, can be small. Thus, $T_{UV}$ is around +10 for $\delta \simeq 0.1$ and around +1 for $\delta \simeq 0.01$. In order to remain in accord with precision electroweak, recall that the net contribution to $T$ from the IR and UV
needs to cancel to a number of order $+0.1$. Returning to equation
(\ref{TUV}), if we literally take $x=1$, then for $\delta\simeq 0.01$, we have
$T_{UV}\approx 1$. So one requires $T_{IR} \approx -1$, implying an order $10\%$  tuning in the Higgs
masses to fit within the $68\%$ confidence ellipse\footnote{This is, of course, the naive tuning. There is no objective measure of tuning.}. On the other hand,  for $\delta\simeq 0.1$, one requires an order $1\%$ tuning.
Note that the lower range of $\delta$ of order $0.01$
also requires less tuning in the quark Yukawas, and leads to a better fit with
precision gauge coupling unification. Perhaps this points to small $\delta$
being favored when there is no custodial $SU(2)$.

Consider next the contribution to $S_{UV}$. In explicit realizations of technicolor theories,
the contribution to $S_{UV}$, based on extrapolation of QCD-like behavior, is
a significant hurdle to overcome. However, this is not true for the theories under consideration.
In addition to being supersymmetric and conformal above $\Lambda_{(0)}$, the
strongly coupled theories which we consider are quite different from QCD (and SQCD). Thus, $S$ does not scale like that in a traditional
technicolor theory. Another important feature is that some of the states of the hidden sector will get masses
which are roughly $SU(2)_L$ preserving. In the MSSM, the Higgs sector is vector-like with respect to
$SU(2)_{L}\times U(1)_{Y}$ and  the contributions from visible
superpartners are negligible since supersymmetry breaking gives them
$SU(2)_{L}$ preserving masses. Similar considerations apply in the
quasi-hidden sectors considered in the present framework. Indeed, if supersymmetry breaking
still gives the dominant contribution to the scalars of the hidden sector, they will have largely
$SU(2)_{L}$ preserving masses.

The fermionic states, however, are assumed to primarily get their mass
from the Higgs vev, so one might worry about the size of such contributions. It is hard to determine $S_{UV}$ from first principles, but we will argue that it can naturally be small enough. If we just view the extra sector states as weakly coupled additional vector-like generations for simplicity, the contribution to the $S$ parameter can be estimated to be $c_{extra} \times 1 / 6 \pi$, where $c_{extra}$ is a coefficient which depends on the representations under the electroweak gauge group as well as the details of Yukawa couplings (see for example \cite{Peskin:1990zt, He:2001tp, Kribs:2007nz}).
For example, for the states appearing in (\ref{4thgen}), one gets:
\begin{equation}
c_{extra} = 8-\log\left(\frac{\kappa_{QU}\tilde{\kappa}_{QD}\kappa_{LE}\tilde{\kappa}_{LS}}{\tilde{\kappa}_{QU}\kappa_{QD}\tilde{\kappa}_{LE}\kappa_{LS}}\right)^2
\end{equation}
Hence a slight hierarchy between the Yukawa couplings $\kappa$ and $\tilde{\kappa}$ can give rise to $|c_{extra}| \lesssim 1$. In models with a single Higgs doublet\footnote{such as usual technicolor models where the role of the Higgs doublet is played by a fermion condensate.}, the difference between the Yukawa couplings $\kappa$ and $\tilde{\kappa}$ above leads to a large positive contribution to $T$ which is a significant problem. However, since we have a two Higgs doublet model, this positive contribution can be cancelled by a negative contribution from $T_{IR}^{\text{doublet}}$ as explained earlier.

Of course, one should not forget that viewing the extra states as weakly coupled vector-like generations description is not quite correct.
In \cite{Dugan:1991ck} it was found that the representation content of the extra states can affect the sign of $S$. Considering that whole towers of states with different electroweak representations contribute in the quasi-hidden CFT, this provides an indication that contributions from different representations can at least partially cancel each other, and either sign contributions to $S_{UV}$ with $|c_{extra}| \lesssim 1$ may  be possible\footnote{As a related comment, though not strictly
relevant for our present discussion, it is interesting to note that both the
magnitude and sign of the analogue of $S$ can be freely adjusted in
supersymmetric $U(1)\times U(1)$ gauge theories. Indeed, viewing $S$ as the
size of the kinetic mixing between these two $U(1)$ factors, we observe that
there is a $2\times2$ matrix of holomorphic gauge couplings $\tau_{ij}$. In
$\mathcal{N}=2$ supersymmetric gauge theories, this matrix of couplings
corresponds to the modular parameters of a genus two Riemann surface. In the
special case of a genus two Riemann surface, the only requirement is that
$\tau_{ij}$ is positive definite. In the perturbative regime,
$\operatorname{Im}\tau_{11}$ and $\operatorname{Im}\tau_{22}$ are both large
and positive. Hence, either sign of $\operatorname{Im}\tau_{12}$ is in
principle possible. It would be interesting to incorporate this observation
into explicit models.}. In addition, once conformal symmetry is broken in the hidden sector and electroweak symmetry is broken in the visible sector, both Dirac-type and Majorana-type (if allowed by gauge symmetry) mass terms for these sates are generated. For example, given a vector like pair $L^{(4)} \oplus \widetilde{L}^{4}$, the higher dimension operator $H_u H_d L^{(4)} \widetilde{L}^{(4)} / M_{\cancel{CFT}}$ can generate a significant $SU(2)_L$ preserving mass for these states at strong coupling, which will lower their contribution to $S_{UV}$. The presence of Majorana mass terms for some of the extra states can give rise to negative contributions to $S_{UV}$ \cite{Gates:1991uu}, which can also partially cancel positive contributions from other states. Thus, combining all contributions $S \equiv S_{UV} + S_{IR}^{\text{singlet}} +S_{IR}^{\text{doublet}}$ can naturally lie between -0.1 and 0.1.

We now consider models in which the quasi-hidden sector enjoys an approximate $SU(2)_{R}$ flavor symmetry.
This can occur in the present class of theories, with $SU(2)_{R}$ appearing as
an emergent flavor symmetry in the infrared, and with the Higgs fields
transforming in the $(2,2)$ of $SU(2)_{L}\times SU(2)_{R}$.\footnote{An
explicit example of this type is the \textquotedblleft$S_{3}$
monodromy\textquotedblright\ scenario of \cite{HVW}.} Note that in the deep
IR, the Higgs fields would have the same dimension $\Delta$. In this case, a
diagonal $SU(2)_{diag}\subset SU(2)_{L}\times SU(2)_{R}$ functions as
custodial $SU(2)$ when $\tan\beta=1$, and deviations from this value
correspond to an additional order parameter, $\varepsilon\simeq\tan\beta-1$
(though even for $\tan\beta=2$ there is some suppression). This leads to an
extra suppression of the parameter $x$ in equation (\ref{TUV}) so that
$T_{UV}\simeq\alpha^{-1}\times\varepsilon^{2}\times\delta$ (see e.g.
equation (\ref{tandisc})). Hence, even with $\delta\simeq\varepsilon\simeq
0.1$, $T_{UV}\simeq 0.1$ can be obtained quite naturally leading to consistency with precision electroweak.

Such small values of $\tan\beta$ are a well-motivated possibility. In the
DSSM, this can arise if $m_u^2 = m_d^2$. If $m_{u(0)}^{2}=m_{u(0)}^{2}$ at the messenger scale, and provided the mediation
scale is sufficiently low, there will not be sufficient RG\ time to distort
this relation by a large amount.\footnote{For example, this could happen in gauge mediation with a low messenger scale.}

\section{Collider Signatures \label{sec:collider}}

In this section we comment on the collider signatures of the framework. The
collider phenomenology is quite rich, though also very dependent on details of
the hidden sector. Consequently we restrict ourselves to making qualitative and
general remarks which arise via the main features of the framework.

Recall that the main features of the DSSM are a modified Higgs sector, and an
accompanying set of extra states, with respective masses:%
\begin{equation}
M_{H}\simeq\sqrt{\delta}\times\Lambda_{soft}\text{, }M_{extra}\simeq
\sqrt{\delta}\times\Lambda_{(0)}.
\end{equation}
These extra states transform in $SU(5)_{GUT}$ multiplets, and so have
$G_{SM}$ quantum numbers. Since $\Lambda_{(0)}\simeq4\pi v\simeq3$ TeV, we see
that in the range $0.01\leq\delta\leq0.1$, this yields a mass scale roughly of
order $300$ GeV $\lesssim M_{extra}\lesssim950$ GeV. The precise mass,
however, depends on order one Yukawa couplings to the hidden
sector. This gives an order one uncertainty to the above estimates. Thus there is a
rich hidden sector with \textquotedblleft mesons\textquotedblright\ which are
either charged or uncharged under $G_{SM}$. The main portal which connects
the visible and hidden sector is the F-term deformation $H_{u}\cO_{u}%
+H_{d}\cO_{d}$.\footnote{In the D3-brane theories of \cite{Funparticles, HVW}
there are additional mixing terms via the third generation, but to keep our
discussion streamlined, we focus on just the minimal interaction term.}

The fact that there is an approximate conformal sector consisting of states
with non-trivial scaling dimensions coupled to the visible sector might at
first suggest a scenario similar to unparticles \cite{Georgi:2007ek}. However, in the DSSM the conformal symmetry is broken at
the scale $\Lambda_{soft}$, which sets the scale for the soft masses around a
TeV. Hence this framework should be viewed as a particular class of hidden valley models, namely
SM charged unparticles with a mass gap \cite{Strassler:2006im, Strassler:2008bv}. In
addition, the Higgs sector can in principle mix with operators in the
conformal sector. Such interactions can have important implications
for Higgs physics, as discussed for example in \cite{Strassler:2008bv, Stancato:2008mp, Stancato:2010ay}. See
\cite{Cacciapaglia:2007jq, Cacciapaglia:2008ns, Cai:2009ax, Cai:2011ww} for related work on the
phenomenology of unparticle scenarios with a mass gap.

To look at the phenomenology in a bit more detail, we divide the discussion
into two parts. By assumption, the states of the hidden sector fill out
$SU(5)_{GUT}$ multiplets. Hence, we first discuss the phenomenology of the
Higgs sector and the operators in the conformal sector charged under just
$SU(2)_{L}\times U(1)_{Y}$, but neutral under $SU(3)_{C}$. Next, we comment on
the additional states charged under $SU(3)_{C}$.

\subsection{Electroweak Sector}

Let us first make some general comments on the signatures of states charged
under $SU(2)_{L}\times U(1)_{Y}$ but which are neutral under $SU(3)_{C}$. This
includes the Higgs sector states, but also states of the hidden sector. The
two sectors mix via couplings such as $H_{u}\cO_{u}+H_{d}\cO_{d}$, which can in
principle have an important consequence for Higgs physics.

Up to now, we have implicitly assumed that the mixing between the
Higgs and hidden sector states is small enough so that the Higgs fields
retain their character. In a weakly coupled
approximation (c.f. equation (\ref{4thgen})), one can view $\cO_{u}$ as containing
$L^{(4)}S_{u}$, where $L^{(4)}$ has quantum numbers
conjugate to $H_{u}$ and $S_{u}$ is a $G_{SM}$ gauge singlet. If $S_{u}$ then
gets a vev, $H_{u}$ and $L^{(4)}$ would also mix.  It is natural to expect some of the
singlets of the hidden sector to get a vev which gives the
$U(1)_{hid}$ gauge boson a mass, as can occur in the D3-brane theories of
\cite{Funparticles, HVW}. The possibility of large mixing with this hidden
sector is quite interesting, but more difficult to analyze.

The case of small mixing is also well-motivated theoretically, and lends itself to
an easier analysis. Small mixing can occur if a hidden sector singlet
$S_{hid}$ with excess dimension $\delta_{hid}$ experiences a squeezed Mexican
hat potential so that the vev is $v_{hid}\simeq\Lambda_{soft}%
\times\left(  q_{hid}\right)  ^{1/2\delta_{hid}}$. Hence, even if the
SM\ singlets of the hidden sector pick up vevs, this scale may be far below
the Higgs vev, thus limiting the mixing between the Higgs and the extra
$SU(2)_{L}\times U(1)_{Y}$ charged states.

Even in the case of weak mixing, the resulting phenomenology can be quite
interesting as it can give rise to new production and decay modes of the Higgs.
The main ideas can be conveyed by considering mesonic operators
$\cO_{a}\sim\phi_{a}^{\dag}\phi_{a}$ of the hidden sector, where the $\phi_{a}$
can refer to states with or without $SU(2)_{L}\times U(1)_{Y}$ quantum
numbers.\footnote{This strictly only makes sense if it is possible to view
$\mathcal{O}_{a}$ as bound state of $\phi_{a}$ and $\phi_{a}^{\dag}$.}
Production of an operator $\mathcal{O}_{a}\simeq\phi_{a}^{\dag}\,\phi_{a}$
through gluon fusion could proceed followed by the decay $\phi^{a}%
\rightarrow\phi^{b}\,H$ if kinematically allowed, resulting in a channel such
as $b\bar{b}\tau^{+}\tau^{-}$ + missing energy if $\phi^b$ decays to Standard Model neutral 
hidden sector states which are either stable or long-lived on collider time scales. On the other hand, for
$m_{H}>2\,m_{\phi_{a}}$, vector boson fusion or $WH/ZH$ associated production
followed by decays $H\rightarrow\phi_{a}\,\phi_{a}$, $\phi_{a}\rightarrow
H^{\ast}\,\phi_{b}$ could result in forward jets and soft jets or leptons and
missing energy.

Just as in other two Higgs doublet models, the lightest Higgs could have a different
branching fraction to gauge bosons, as well as other Standard Model states. In fact, if it
is sufficiently heavy, it may also decay to $t \overline{t}$. In addition to these more
``standard'' decay modes, there is the possibility that
the Higgs can decay to hidden sector states or LSPs, reducing the branching ratio to visible sector states. Its branching ratio
to two gluons and two photons will also be generically different than in the MSSM because of the presence of extra scalar and fermionic states which run inside the loop.  Hence, the latest Higgs bounds from the LHC are generically weakened when interpreted within this framework. Of course, the precise form of the resulting signatures is model-dependent. At
one extreme, if the extra states are heavier than the Higgs by a factor of
$\Lambda_{(0)}/\Lambda_{soft}$, the main Higgs decays to the hidden sector
involve higher dimension operators, which would be suppressed relative to
decays to SM states. However, decays to LSPs, if the LSP is in the visible sector, may still occur
with a non-trivial branching ratio. At the other extreme, if there is a significant mixing between
the Higgs and the hidden sector, the resulting phenomenology will be quite
rich and complicated. This could give rise to a scenario with multiple
cascade decays of the Higgs (or the operators it mixes with) and high
multiplicity in the final state. These kinds of signatures have been
considered in \cite{Strassler:2006im, Han:2007ae, Strassler:2008bv} from a
phenomenological perspective. We can therefore view the above framework as
providing an explicit UV-motivated realization of such scenarios.

\subsection{Colored Sector}

The DSSM\ also naturally contains states charged under $SU(3)_{C}$, which are
necessary to retain gauge coupling unification. Given the significant
improvement in search channels for light colored states coming from ATLAS and
CMS, it is important to study the consequences of these searches for the
present class of models.

The states of the hidden sector can decay when there are additional Yukawa couplings to the visible sector. This is
what occurs, for example, in the string motivated scenarios of \cite{Funparticles}, where the third generation states couple
to additional hidden sector operators. In many cases, these couplings are irrelevant, and so if we had flowed
to the deep IR, would have vanished. However, since we have cutoff of the RG flow, they contribute an additional
source of couplings which can allow colored (and uncolored) states of the hidden sector to decay. The decay
of such states in related scenarios has been studied in \cite{D3gen}. For CFT breaking near the weak scale,
the decay rate is on the order of $\Gamma \sim M_{weak} (M_{weak} / M_{GUT})^{\nu}$ for $\nu$ a small parameter
related to the excess dimension of the third generation operators. This leads to decays which are prompt on the order of
collider timescales.

Let us now discuss some additional features of these extra states. Since
the states in the hidden conformal sector are supersymmetric, the
scalar and fermionic operators have different $R$-parity
assignments.\footnote{We assume for concreteness that $R$-parity is conserved.
In the particular context of F-theory GUT\ models, matter parity can be viewed
as a discrete subgroup of a $U(1)$ Peccei-Quinn symmetry. This $U(1)_{PQ}$
originates from a flavor, e.g. $7^{\prime}$-brane which intersects the visible
sector GUT $7$-brane. The visible sector states correspond to $7-7^{\prime}$
strings, while the states of the D3-brane theory correspond to $3-7$ strings,
and $3-7^{\prime}$ strings. In other words, the states charged under
$SU(5)_{GUT}$ coming from the D3-brane will have even matter parity, while the
hidden sector states neutral under $SU(5)_{GUT}$ may have either matter
parity.} The $R$-parity even states could be produced singly as a resonance.
These states will eventually decay to visible jets (as required by color flow)
and to hidden sector singlets, which could be much lighter. If these hidden sector singlet 
states are sufficiently heavy, they can also eventually decay back to the visible sector
light quarks and leptons. This could happen for example through connector
operators such as $\mathcal{O}_{u},\mathcal{O}_{d}$ or hidden gauge bosons
which mix with the $Z$. Thus, there could be multiple jets and/or
leptons in the final state, many of which could be soft. Thus
current bounds on such colored states from dijet searches such as
\cite{Collaboration:2011ns} may not apply. If, on the other hand, these 
hidden sector singlet states are sufficiently light, then the decay processes involving a Standard Model charged extra state 
will involve a fewer number of jets with larger missing energy. Even in this case, however, the effects of strong coupling 
can in principle produce additional jets. This will affect the branching ratios of the various decay channels. More study is 
needed to determine the relevant bounds in this situation.

$R$-parity odd colored states, on the other hand, can only be pair-produced.
Again, these could decay to visible jets and hidden non-colored states. The
$R$-parity odd states in the hidden sector will eventually decay to multiple
soft jets/leptons and the LSP. Recent ATLAS and CMS searches based on multiple
jets + missing energy are sensitive to these processes, if there are enough
hard jets to trigger on. In such cases, lower bounds on the masses of such
states could be placed. However, it should be clear that the precise bounds
depend on many model dependent details.

The presence of additional colored states from the hidden sector can also
affect Higgs production. An important process of this type is gluon fusion, with a hidden sector
colored state running in the corresponding loop. The size of this contribution is determined by
the coefficient of the effective operator $(H / M_{extra}) \mathrm{Tr}_{SU(3)} F_{SU(3)}^{2}$. This is in turn controlled by
the one loop beta function contribution from new states, as explained for example in \cite{Ellis:1975ap, Shifman:1979eb}.
At high scales where supersymmetry is restored, the size of this threshold
correction for representative DSSM scenarios has been computed in \cite{FCFT}, which we can interpret as 
the coefficient of the $Hgg$ vertex generated by such corrections. In this limit, the contribution can lead to a sizable
increase in the Higgs production cross section (a factor of roughly $O(1) - O(10)$).
In the presence of supersymmetry breaking effects,
however, the form of such threshold corrections is more subtle. For example, when the
soft masses and mixing terms for the stop sector of the MSSM are comparable in size,
the contribution to the gluon fusion effective vertex from the top and stop
sector can cancel \cite{Djouadi:1998az, Dermisek:2007fi}. Similar considerations apply for colored hidden sector states.
Indeed, the generic expectation is that the soft masses and A-terms of the hidden sector will naturally
be of similar scale. In addition, there can be additional suppression/enhancement of gluon fusion 
due to the dependence on the mixing angles in a general two Higgs doublet sector (see for example \cite{Branco:2011iw} for a review). 
Thus, we can expect that there exist regions of parameter space in the hidden sector where the Higgs
production can be similar to that in the MSSM, or perhaps even reduced. As a final, less motivated possibility, one might also
consider scenarios where the colored extra states receive an additional large (compared to the electroweak scale) contribution to their masses,
leaving only uncolored objects in the low energy spectrum. In such cases, the contribution from the hidden sector to processes such as gluon fusion would almost be the same as in the MSSM. Turning the discussion around, collider constraints on the product of the Higgs production cross section and branching fraction can then impose constraints on the properties of the hidden sector. The basic feature,
however, is that based on the qualitative considerations just discussed, we can expect large regions of parameter space in the hidden
sector to be in accord with present constraints on the Higgs sector.

\section{Conclusions \label{sec:CONCLUDE}}

A natural feature of many supersymmetric theories is a non-trivial K\"{a}hler
potential. In this paper we have investigated the consequences of a class of
singular K\"{a}hler potentials which would naturally arise if the Higgs
interacts with a quasi-hidden sector. We have seen that even in the limit
where the Higgs fields have dimension close to one, gauge boson masses are
naturally lower than the Higgs mass. This situation is quite different from the MSSM.
Moreover, whereas many models with a strongly coupled Higgs also face
significant obstacles with low Landau poles, we have seen that the present
class of models can naturally accommodate precision gauge coupling unification
as well as large quark Yukawas, while still being consistent with electroweak precision constraints.
The singular nature of the K\"{a}hler potential also indicates the presence of additional colored and uncolored
states close to the Higgs mass.

Some potential avenues of future investigation are as follows. Our analysis in this paper has focussed on the low energy consequences of
making the Higgs have non-trivial scaling dimension. It would be quite
interesting to realize a more UV-complete version of these dynamics which
includes both a supersymmetry breaking sector, and means of transmission to
the visible sector. The more general lesson we have arrived at is that a non-trivial K\"{a}hler
potential can lead to novel and phenomenologically viable models of
electroweak symmetry breaking. It would be quite interesting to widen the
scope of our investigation to study the broadest possible K\"{a}hler
potential consistent with present phenomenological constraints.

The DSSM predicts extra states, leading to a potentially rich phenomenology which is
also quite model-dependent. Given the data being accumulated at the LHC, a comprehensive
study of even some representative examples would be extremely interesting.

\section*{Acknowledgements}

We thank N. Arkani-Hamed, T. Cohen, N. Craig, D. Gaiotto, D. Green, P. Langacker, G.
Moore, A. Pierce, D. Poland, E. Ponton, N. Seiberg and Y. Tachikawa for helpful
discussions. We also thank P. Langacker and E. Ponton for helpful comments on the draft.
JJH, PK and BW thank the Harvard high energy theory group for
hospitality during part of this work. JJH, PK, CV\ and BW\ thank the 2011
Simons workshop in Mathematics and Physics and the Simons Center for Geometry
and Physics for hospitality during the completion of this work. The work of
JJH is supported by NSF grant PHY-0969448. The work of PK\ is supported by
DOE\ grant DE-FG02-92ER40699. The work of CV\ is supported by NSF\ grant
PHY-0244821. The work of BW is supported by DOE\ grant DE-FG02-95ER40899.

%%%%%%%%%%%%%%%%%%%%%%%%%%%%%%%%%%%%%%%%%%%%%%%%%%%%%%%%%%%%%%%%%%%%%%%%%%%%%%%%%%%%%%%%%%%%%%%%%%%%%%%%%%%

\appendix

\section{Review of D3-Brane CFTs}\label{d3cft}

In this Appendix we provide a brief review of some of the properties of the
D3-brane CFTs introduced in \cite{Funparticles} and studied further in
\cite{FCFT, TBRANES, D3gen, HVW}.\ Though the natural setting for these CFTs
is from a string construction, many of their properties can be stated in
purely field theory terms. The starting point is a four-dimensional conformal field theory
with $\mathcal{N}=2$ supersymmetry. From these \textquotedblleft master
theories\textquotedblright\ other conformal field theories can be obtained by
performing relevant deformations to new theories with $\mathcal{N}=1$ supersymmetry.

The starting \textquotedblleft master theories\textquotedblright\ are known as
Minahan-Nemeschansky E-type theories \cite{MNI, MNII}. These are
four-dimensional theories with $\mathcal{N}=2$ supersymmetry and a flavor
symmetry which is $E_{6}$, $E_{7}$ or
$E_{8}$. For concreteness, we focus on the $E_{8}$
theory, since the others can be obtained from deformations of this
one. These theories are intriniscally strongly coupled, and there is no known weakly
coupled Lagrangian description available. The lack of such a description
does not mean such theories are inaccessible to study.
Indeed, the absence of a Lagrangian is compensated by the presence of $\mathcal{N}=2$
supersymmetry and the associated Seiberg-Witten curve. The operators of this theory transform in representations
of $E_{8}$. There is a particular class of operators $\cO_{248}$ transforming in
the $248$ (adjoint) of $E_{8}$. These operators have dimension two
and can loosely be thought of as the analogue of the mesons in SQCD.

From this $\mathcal{N}=2$ master theory we obtain the $\mathcal{N}=1$ CFTs
describing the D3-brane probe CFTs of \cite{Funparticles}. We view the SM gauge group as embedded in $E_{8}$ via:%
\begin{equation}
G_{SM}\subset SU(5)_{GUT}\subset SU(5)_{GUT}\times SU(5)_{\bot}\subset E_{8}.
\end{equation}
This is obtained by switching on operator deformations of the $\mathcal{N}=2$
theory involving the dimension two ``mesonic'' operators. Such deformations
break the flavor symmetry $E_{8}$ to $SU(5)_{GUT}$. These
deformations generically also break $\mathcal{N}=2$ to $\mathcal{N}=1$. The
original operators $\cO_{248}$ then decompose into irreducible representations
of $G_{SM}$. In particular, some of these operators will have quantum numbers
conjugate to the Standard Model fields. One therefore expects the operator
deformations:%
\begin{equation}
\int d^{2}\theta\text{ }\Psi_{R} \cO_{\overline{R}}%
\end{equation}
for $\Psi_{R}$ a chiral superfield of the MSSM in representation $R$ of
$G_{SM}$, and $\cO_{\overline{R}}$ an operator in the conjugate representation.
This leads to operator couplings such as:
\begin{equation}
\int d^{2}\theta\text{ }\left ( H_{u} \cO_{u}+H_{d} \cO_{d} \right).
\end{equation}
Let us note that in the stringy realization of these theories, there are typically
additional couplings to the third generation:
\begin{equation}
\int d^{2}\theta\text{ }\Psi_{R}^{(3)} \cO_{\overline{R}}%
\end{equation}
which can also alter the dimension of the third generation fields. We
emphasize that such couplings are simply a part of the UV definition of the
theory, and must be included in a consistent UV treatment. Of course, from the
perspective of the low energy field theory, one can switch off such couplings.
In this paper we have mainly focussed on the simplest realization
of a DSSM where only mixing with the Higgs fields is included.

Including all of these operator deformations, one can use the powerful
technique of a-maximization \cite{Intriligator:2003jj} to determine the
resulting dimensions of the Higgs and other SM fields. A remarkable
feature of many of these infrared fixed points is that the chiral superfields of the
visible sector have scaling dimensions which are close to one. That is, the
visible sector retains its identity, even in the deep infrared.

\section{Explicit Example \label{examples}}

In this Appendix we study the mass spectrum in a situation where the doublets $H_u = (H_u^{+} , H_u^0)$
and $H_d = (H_d^0 , H_d^-)$ both have dimension $\Delta$, and the
K\"{a}hler potential is taken to be:
\begin{equation}
K_{\Delta}=(H_{u}^{\dag}H_{u}+H_{d}^{\dag}H_{d})^{1/\Delta}.
\end{equation}
This choice of K\"{a}hler potential has the additional benefit of preserving
$SU(2)_{L}\times SU(2)_{R}$, where the Higgs doublets transform in the
$(2,2)$. This extra $SU(2)_{R}$ means that custodial $SU(2)$ can also be
preserved near the $\tan\beta\rightarrow1$ limit (see equation
(\ref{tandisc})).

A general discussion of supersymmetric Lagrangians with a non-trivial K\"ahler potential is given in \cite{Bagger:1982fn, Hull:1985pq}, and a review can be found in chapter XXIV of \cite{WESSBAGGER}. Focussing on the Higgs sector, the system of interest consists of a four-component
vector $\Phi^{i} = (H_u^{+} , H_u^{0} , H_{d}^{0} , H_{d}^{-})$ given by the usual two Higgs doublets.
The Lagrangian of the scalar fields is:
\begin{equation}
L_{scalar}=-g_{i\overline{j}}D_{\mu}\Phi^{i}D^{\mu}\Phi^{\dag \bar j}-\frac{1}%
{2}g^{2}D^{(a)}D^{(a)} -\frac{1}{2}g^{\prime 2}D^{2} -g^{i\overline{j}}\partial_{i}W\partial_{\overline{j}%
}W^{\dag}%
\end{equation}
where $W$ is the superpotential, and the auxiliary fields $D^{(a)}$ and $D$ of
the vector multiplet associated with $SU(2)_{L} \times U(1)_{Y}$ are:
\begin{align}
D^{(a)} &  = - \frac{1}{\Delta}\left(  H_{u}^{\dag}\cdot \tau^{(a)}\cdot H_{u}%
+H_{d}^{\dag}\cdot \tau^{(a)}\cdot H_{d}\right)  \times\left(  H_{u}^{\dag}%
H_{u}+H_{d}^{\dag}H_{d}\right)  ^{(1-\Delta)/\Delta}\\
D &  = - \frac{1}{2\Delta}\left(  H_{u}^{\dag}H_{u}-H_{d}^{\dag}H_{d}\right)
\times\left(  H_{u}^{\dag}H_{u}+H_{d}^{\dag}H_{d}\right)  ^{(1-\Delta)/\Delta}%
\end{align}
where the $\tau^{(a)}$'s are spin 1/2 generators of $SU(2)$.

\begin{figure}[t]
\begin{center}
\includegraphics[
height=2.5564in,
width=3.9885in
]{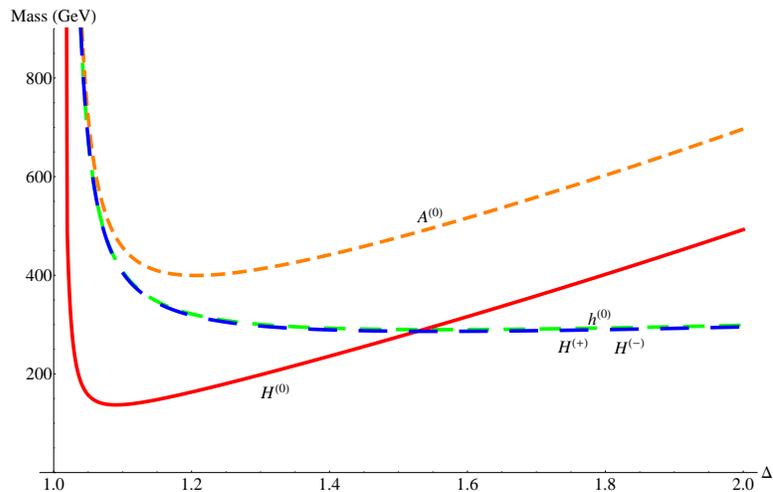}
\end{center}
\caption{Plot of Higgs boson masses for a representative scenario with
$\tan\beta=1$. The numerical values used in the plot are $B_{0}=1$, $m_{0}%
^{2}=0$ and $q_{0}=0.9$. In these plots, the mass of the $Z$-boson has been
held fixed at its observed value. By inspection, we see that as $\Delta
\rightarrow1$, the resulting masses diverge. Observe that at $\Delta$ near
one, $H^{(0)}$ is the lightest neutral Higgs, while near two, $h^{(0)}$ is the
lightest neutral Higgs.}%
\label{newnewfirstplot}%
\end{figure}

\begin{figure}[t]
\begin{center}
\includegraphics[
height=2.5564in,
width=3.9885in
]{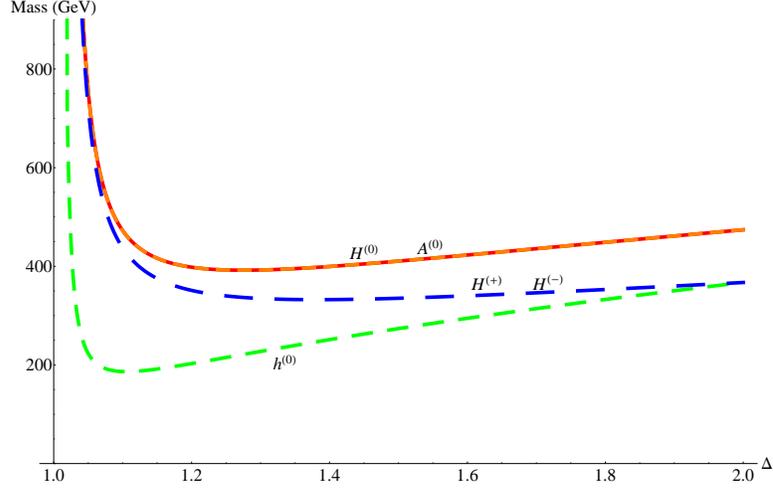}
\end{center}
\caption{Plot of Higgs boson masses for a representative scenario with
$\tan\beta=\infty$. The numerical values used in the plot are $B_{0}=0$,
$m_{u(0)}^{2}=-1$, $m_{d(0)}^{2}=+1$ and $q_{0}=0.9$. In these plots, the mass
of the $Z$-boson has been held fixed at its observed value. By inspection, we
see that as $\Delta\rightarrow1$, the resulting masses diverge. Observe that
in this case $h^{(0)}$ is always the lightest Higgs.}
\label{tanbbigplot}%
\end{figure}

In the remainder of this section we study two particular scenarios, one with
$\tan\beta=1$ and one with $\tan\beta=\infty$. Our notation for the physical Higgs
spectrum is as in \cite{MartinPrimer}, e.g. $(h^{0}, H^{0}, A^{0}, H^{+},
H^{-})$, with corresponding mixing angles $\alpha$, $\beta_{\pm}$, $\beta_{0}%
$. See figures \ref{newnewfirstplot} and \ref{tanbbigplot} for plots of the
mass spectra in these cases. The rest of the supersymmetric Higgs sector, in
particular the mixing with gauginos in the chargino and neutralino system, can be
determined using the general expressions given in \cite{WESSBAGGER}. The end result
is not particularly illuminating, so we suppress it in what follows.

\subsection{$\tan\beta=1$}

In this section we study a choice of parameters which leads to electroweak
symmetry breaking with $\tan\beta=1$. To realize this case, we take
$m_{u}^{2}=m_{d}^{2}$. In the weakly coupled MSSM, the limit $\tan
\beta\rightarrow1$ does not admit a metastable vacuum which both breaks
electroweak symmetry and is bounded from below. Here we show that such critical
points do exist in the DSSM, and we study the resulting mass spectra. To further simplify our analysis, we also take
the soft masses to vanish. None of our qualitative conclusions depend on this requirement.

In this case, the Higgs vev in this symmetry breaking minimum are:
\begin{equation}
v_{u}^{2} = v_{d}^{2} = \frac{1}{2} \Lambda_{soft}^{2 \Delta} \times \left(  q_{0}\right)  ^{\Delta/(\Delta-1)}.
\end{equation}
Here, the parameter $q_{0}$ is a dimensionless ratio of the parameters of the
theory:%
\begin{equation}
q_{0}=\frac{B_{0} / \mu_{0}^{2}}{2\Delta^{2}-\Delta}.
\end{equation}
Consistency of the approximation we are considering requires the dimensionless
ratio $q_{0}<1$. Let us note that if we view the soft breaking parameters as
radiatively generated, and in particular, smaller than the supersymmetry
preserving term $\mu_{0}$, this type of situation can be arranged.\footnote{As
a simple example in the context of the MSSM, we note that if the $\mu$-term is
generated at some messenger scale but the $B\mu$-term vanishes at this
boundary condition, $B\mu$ will be radiatively generated, and naturally
suppressed relative to $\mu^{2}$. However, it can still be larger in magnitude than the
soft masses squared. Finally, we note that in gauge mediation scenarios,
the requirement $m_u^2 = m_d^2$ at the messenger scale is also easily arranged. Provided the messenger
scale is not too high, this leads to similar Higgs potentials to those we study here.}

Let us now summarize the mass spectrum of the various states which are in
direct contact with our Higgs operators. The gauge boson masses squared are:%
\begin{equation}
\frac{M_{W,cl}^{2}}{\Lambda_{soft}^{2}}=\frac{g^{2}}{2\Delta}\times
(q_{0})^{1/(\Delta-1)}\text{, }\frac{M_{Z,cl}^{2}}{\Lambda_{soft}^{2}}%
=\frac{g^{2}+g^{\prime2}}{2\Delta}\times(q_{0})^{1/(\Delta-1)}\text{.}%
\end{equation}
The parameter $\Lambda_{soft}$ which defines the scale of soft supersymmetry
breaking parameters in the Higgs potential is related to the Z-boson mass
squared:%
\begin{equation}
\frac{\Lambda_{soft}^{2}}{M_{Z,cl}^{2}}=\frac{2\Delta}{g^{2}+g^{\prime2}%
}\left(  q_{0}\right)  ^{-1/(\Delta-1)}.
\end{equation}
The Higgs mixing angles are:
\begin{equation}
\cos\beta_{0}=\sin\beta_{0}=\cos\beta_{\pm}=\sin\beta_{\pm}=\cos\alpha
=\sin\alpha=1/\sqrt{2}.
\end{equation}
The Higgs masses squared are:
\begin{align}
\frac{M_{h^{(0)}}^{2}}{M_{Z}^{2}}  &  =\frac{4\Delta^{3}\times\left(
q_{0}\right)  ^{(\Delta-2)/(\Delta-1)}}{g^{2}+g^{\prime2}}\times\frac
{1}{\Delta}\times\left(  \left(  \frac{1}{2\Delta-1}\right)  B_{0}+\frac
{g^{2}+g^{\prime2}}{4\Delta^{2}}\times\left(  q_{0}\right)  ^{(2-\Delta
)/(\Delta-1)}\right) \\
\frac{M_{H^{(0)}}^{2}}{M_{Z}^{2}}  &  =\frac{4\Delta^{3}\times\left(
q_{0}\right)  ^{(\Delta-2)/(\Delta-1)}}{g^{2}+g^{\prime2}}\times\left(
\frac{\Delta-1}{\Delta}\right)  \times B_{0}\\
\frac{M_{A^{(0)}}^{2}}{M_{Z}^{2}}  &  =\frac{4\Delta^{3}\times\left(
q_{0}\right)  ^{(\Delta-2)/(\Delta-1)}}{g^{2}+g^{\prime2}}\times B_{0}\\
\frac{M_{H^{(\pm)}}^{2}}{M_{Z}^{2}}  &  =\frac{4\Delta^{3}\times\left(
q_{0}\right)  ^{(\Delta-2)/(\Delta-1)}}{g^{2}+g^{\prime2}}\times\frac
{1}{\Delta}\times\left(  \left(  \frac{1}{2\Delta-1}\right)  B_{0}+\frac
{g^{2}}{4\Delta^{2}}\times\left(  q_{0}\right)  ^{(2-\Delta)/(\Delta
-1)}\right)  .
\end{align}
From this behavior, we see that $\left(  H^{(+)},h^{(0)},H^{(-)}\right)  $
form an approximate triplet of custodial $SU(2)$, with masses of the same size
as $\Lambda_{soft}$. Moreover, near $\Delta\rightarrow1$, $H^{(0)}$ is the
lightest Higgs, while in the other limit where $\Delta\rightarrow2$, $h^{(0)}$
is the lightest Higgs. Observe also that the relative mass of the Higgs bosons
to the gauge boson masses is quite sensitive to the value of the parameter
$q_{0}$, and in particular, the overall factor of $\left(  q_{0}\right)
^{(\Delta-2)/(\Delta-1)}$. Since $1<\Delta<2$, observe that for $q_{0}<1$, the
Higgs is parametrically heavier. One should exercise some caution because this
is really a tree level analysis; as we push the Higgs mass above $800$ GeV,
gauge boson scattering ceases to be perturbative.

\subsection{$\tan\beta=\infty$}

We now study the Higgs masses in the opposite limit where $B=0$ and we take a
particular limit where the up-type soft masses are tachyonic, while the down-type
soft masses are regular. This leads to $\tan\beta=\infty$, namely $H_{u}$
alone gets a vev. Similar symmetry breaking patterns can be arranged in the MSSM.
To simplify the analysis further, we drop the contribution from
the D-terms, as it is a very small change in the form of the potential.

With this approximation scheme, the Higgs vevs satisfy:%
\begin{equation}
v_{u}^{2}=\Lambda_{soft}^{2\Delta}\times\left(  q_{0}\right)  ^{\Delta
/(\Delta-1)}\text{, }v_{d}^{2}=0
\end{equation}
where the dimensionless constant $q_{0}$ is:%
\begin{equation}
q_{0}=\left(  \frac{-m_{u(0)}^{2}/\mu_{(0)}^{2}}{2\Delta^{4}%
-\Delta^{3}  }\right)  ^{1/2}.
\end{equation}
Note that in order to achieve a symmetry breaking vev, $m_{u(0)}^{2}<0$. The
gauge boson masses squared are:%
\begin{equation}
\frac{M_{W,cl}^{2}}{\Lambda_{soft}^{2}}=\frac{g^{2}}{2\Delta}\times
(q_{0})^{1/(\Delta-1)}\text{, }\frac{M_{Z,cl}^{2}}{\Lambda_{soft}^{2}}=\frac
{g^{2}+g^{\prime2}}{2\Delta^{2}}\times(q_{0})^{1/(\Delta-1)}\text{.}%
\end{equation}
Observe that in this case, $\rho_{cl}$ is proportional to $\Delta$.

The analysis of Goldstone modes in this situation is actually much simpler. In
this limit we have $\cos\beta_{0}=\cos\beta_{+}=0$, and the two physically
equivalent branches $\sin\alpha=0$ and $\cos\alpha=0$. We specify the form of
the spectrum for the choice of mixing angles:%
\begin{equation}
\cos\beta_{0}=\cos\beta_{+}=\sin\alpha=0.
\end{equation}
The resulting Higgs masses squared, are, in units of the $Z$-boson mass:%
\begin{align}
\frac{M_{h^{(0)}}^{2}}{M_{Z}^{2}}  &  =\frac{\left(  q_{0}\right)
^{-1/(\Delta-1)}}{g^{2}+g^{\prime2}}\times(\Delta-1)\times(-8m_{u(0)}^{2})\\
\frac{M_{H^{(0)}}^{2}}{M_{Z}^{2}}  &  =\frac{\left(  q_{0}\right)
^{-1/(\Delta-1)}}{g^{2}+g^{\prime2}}\times\frac{2\Delta}{2\Delta-1}%
\times(-m_{u(0)}^{2}(6\Delta-5)+m_{d(0)}^{2}(2\Delta-1))\\
\frac{M_{A^{(0)}}^{2}}{M_{Z}^{2}}  &  =\frac{\left(  q_{0}\right)
^{-1/(\Delta-1)}}{g^{2}+g^{\prime2}}\times\frac{2\Delta}{2\Delta-1}%
\times(-m_{u(0)}^{2}(6\Delta-5)+m_{d(0)}^{2}(2\Delta-1))\\
\frac{M_{H^{(\pm)}}^{2}}{M_{Z}^{2}}  &  =\frac{\left(  q_{0}\right)
^{-1/(\Delta-1)}}{g^{2}+g^{\prime2}}\times2\Delta\times\left(  -m_{u(0)}%
^{2}+m_{d(0)}^{2}\right)
\end{align}
We observe that $h^{(0)}$ is the lightest Higgs. Next lightest are the charged
Higgs fields $H^{(+)}$ and $H^{(-)}$. Finally, the heaviest Higgs bosons are
$A^{(0)}$ and $H^{(0)}$, which are degenerate at this level of approximation.
See figure \ref{tanbbigplot} for a plot of these masses as a function of
$\Delta$.

\bibliographystyle{utphys}
\bibliography{ConfHiggs}

\end{document}